\documentclass[12pt]{iopart}
\usepackage{amssymb}
\usepackage{graphicx,subfigure}
\usepackage[utf8]{inputenc}
\begin{document}

\title{Thermodynamics and structure of simple liquids in the hyperbolic plane}
\author{Fran\c{c}ois Sausset,  Gilles Tarjus and Pascal Viot}

\address{Laboratoire de Physique Th\'eorique de la Mati\`ere Condensée,
 Université Pierre et Marie Curie-Paris 6, UMR CNRS 7600, 4 place Jussieu,
 75252 Paris Cedex 05, France}

\ead{sausset@lptmc.jussieu.fr,tarjus@lptmc.jussieu.fr,viot@lptmc.jussieu.fr}
\begin{abstract}
We  provide   a  consistent   statistical-mechanical  treatment   for
describing the thermodynamics and  the structure of fluids embedded in
the hyperbolic plane. In particular, we derive a generalization of the
virial  equation relating the  bulk thermodynamic pressure to the pair
correlation  function and    we develop the   appropriate setting  for
extending the  integral-equation approach  of  liquid-state  theory in
order  to  describe the fluid  structure. We  apply the  formalism and
study the   influence  of negative space  curvature   on two  types of
systems that have been recently considered: Coulombic systems, such as
the  one-  and  two-component  plasma  models, and  fluids interacting
through short-range pair potentials,   such as the hard-disk  and  the
Lennard-Jones models.
\end{abstract}

\maketitle

\section{Introduction}\label{sec:introduction}

When  a  system  is embedded  in  a space   of nonzero curvature,  its
long-distance properties, such as the thermodynamic quantities and the
critical   behavior, are  strongly   modified.   A   uniform  positive
curvature, characteristic  of spherical geometry,  has the drawback of
leading to spaces of finite  extent. Genuine long-distance  properties
on the other hand are accessible in hyperbolic spaces characterized by
a constant negative curvature. The influence of the negative curvature
of  hyperbolic space on critical  phenomena has been investigated both
for         hard        spins             on                hyperbolic
lattices\cite{Rietman1992,Wu1996,Wu2000,Angl`esd'Auriac2001,Benjamini2001,Shima2006,Shima2006a,Ueda2007,Baek2007,Gendiar2008,Baek2008b,krcmar:061119,baek:011124}
and   in           a              continuum          field-theoretical
description\cite{Callan1990,Benjamini2001,Doyon2004a,Belo2007}.

In brief, the presence   of an intrinsic characteristic
length associated with  the   metric makes the critical  behavior   of
Ising-type systems mean-field-like\cite{Rietman1992,Wu1996,Angl`esd'Auriac2001,Doyon2004a,Shima2006,Ueda2007}
and drastically alters, if  not suppresses, critical points in  models
with   a   continuous symmetry\cite{Callan1990,Belo2007,Baek2007,Baek2008b}
In  the   last decade,  the  thermodynamics  of simple fluids   in the
hyperbolic plane has also received special attention. This is the case
for   systems interacting through the  Coulomb  potential, such as the
one-              and             two-component                 plasma
models\cite{jancovici2004,fantoni2003,Jancovici1998}    and   for  the
hard-disk fluid\cite{modes:235701,modes:041125,haro:116101}.

In addition to being   a mathematical curiosity and providing   simple
systems for studying the influence of space curvature on long-distance
properties, fluids in the hyperbolic plane have been considered in the
context of  jamming   phenomena and  glass formation  in  liquids. The
rationale   behind   this  comes  from   the    concept of ``geometric
frustration'' that  describes an incompatibility between the preferred
local      order    and       the   tiling        of    the      whole
space\cite{Sadoc:1999,Nelson:2002}.
  Geometric  frustration     has
emerged in the theoretical description of glasses and amorphous solids
from    the consideration  of  local  icosahedral   order in  metallic
glasses.  No  such frustration  however   operates  in assemblies   of
monodisperse     spherical   particles     in   ordinary    (``flat'')
two-dimensional            space,          and     Nelson          and
coworkers\cite{Nelson:1983a,PhysRevB.28.6377,Nelson:2002}
have proposed  to  curve two-dimensional  space  in order to introduce
frustration and thereby  mimic glasses in three-dimensional space. Two
recent investigations  have been carried  out along these lines. Modes
and  Kamien\cite{modes:235701,modes:041125} have  focused on isostatic
packings  in hyperbolic space,    the isostatic property  having  been
argued  to play  an important role   in  the physics  of nearly jammed
random packings of particles in Euclidean three-dimensional space\cite{Brito2006b}. On the other hand, building on a frustration-based approach of
the glass  transition\cite{Kivelson:1995,Tarjus:2005},  we  have  studied  by
molecular dynamics simulation the  slowing down  of the dynamics  with
decreasing  temperature    in  a   monatomic  liquid:   a  truncated
Lennard-Jones model in the hyperbolic plane\cite{sausset:155701}.

In the  present  work, we provide  a consistent statistical-mechanical
treatment for  describing the  thermodynamics   and the  structure  of
fluids embedded in  the hyperbolic plane.   The paper is  organized as
follows.  In  section   \ref{sec:gener-viri-equat}, we   introduce the
statistical-mechanical framework for  systems in the hyperbolic plane.
By using a variant of the Green-Bogoliubov  method, we relate the bulk
thermodynamic pressure to the pair  distribution function. In  section
\ref{sec:integral-equations},  we   extend  the      integral-equation
approach, which is commonly used  in liquid-state theory, to calculate
the pair correlation   function in  hyperbolic geometry.   We  discuss
thermodynamic   consistency and   gas-liquid   critical behavior.   In
section  \ref{sec:coulombic-systems}, we focus  on Coulombic  systems,
such as   the  one- and two-component   plasma models.   We  apply the
formalism developed in  the previous sections  to two cases for  which
the  pair correlation functions are known  exactly:  the limit of high
temperature and   a  special value of   the  temperature.   In section
\ref{sec:short-range-potent}, we   present the numerical  solution  of
standard   approximate integral   equations  of  liquid-state  physics
(namely, the Percus-Yevick and  hypernetted chain equations) that have
been obtained   after making  use  of  the  harmonic  analysis  in the
hyperbolic plane.  We consider  the hard-disk fluid and the  (truncated)
Lennard-Jones  liquid   and  compare with   existing simulation  data.
Finally,  section  \ref{sec:conclusion} contains our   concluding   remarks,  and some
technical details are provided in two appendices.

\section{The generalized virial equation of state}\label{sec:gener-viri-equat}

For  an atomic liquid at equilibrium  in  the grand-canonical ensemble
(fixed temperature $T$, total  area $A$ and chemical potential $\mu$),
the partition function is given by
\begin{equation}\label{eq:GC}
\Xi(T,A,\mu)=\sum_{N=1}^{\infty}\frac{(Z_1 e^{\beta \mu})^N}{N!A^N}
\int \mathrm{d}{\bf A}^N \exp(-\beta \sum_{i=1}^{N}\sum_{j>i}u(r_{ij}))
\end{equation}
where  $u(r)$  is  the   pair   potential with  $r$   the  appropriate
interatomic geodesic   distance  (for simplicity,   we  only  consider
pairwise additive interactions),  $\beta=1/(k_B T)$, and $Z_1(T,A)$ is
the partition function  of a single  particle, \textit{i.e.} the ideal
gas  contribution.     The   above   expression  is  valid    for  any
two-dimensional space with  $dA$ the appropriate infinitesimal surface
area. It  therefore  applies as  well to  the   hyperbolic plane, with
however some subtelties that are discussed below.

The hyperbolic plane $H^2$ is  a two-dimensional manifold of  constant
negative curvature, whose metric  in polar  coodinates $(r, \phi)$  is
given by
\begin{equation}\label{eq:1}
{\mathrm{d}}s^2= \mathrm{d}r^2 + \left(\frac{\sinh(\kappa  r)}{\kappa}
\right)^2 {\mathrm{d}}\phi^2.
\end{equation}
The  Gaussian   curvature  of $H^2$    is  $-\kappa^2$. The  curvature
parameter $\kappa > 0$ therefore measures  the deviation from ``flat''
(Euclidean) space;  by an abuse  of language, we shall sometimes refer
to $\kappa^{-1}$ as  the ``radius of  curvature''. (Note that $H^2$ is
not embeddable in  three-dimensional Euclidean  space and  is  usually
described through representations  such as  the Poincar\'e disk  model
that is   a conformal   projection of  $H^2$  onto   a disk of  radius
$\kappa^{-1}$.)   In  polar  coordinates,   the  differential  area is
expressed as
\begin{equation}\label{eq:2}
dA=\frac{\sinh(\kappa r)}{\kappa}dr d\phi.
\end{equation}
The  hyperbolic plane  is   of infinite  extent  but the   exponential
character of  the metric at  large distance (see Eqs. (\ref{eq:1}) and
(\ref{eq:2})) induces the peculiar   property that the boundary of   a
finite region of $H^2$ grows as fast as the total  area of this region
when the latter increases. A simple illustration is provided by a disk
of radius $r$: from Eq. (\ref{eq:2}), one finds that its area is equal
to $A(r)=2\pi (\cosh(\kappa  r) -1)/\kappa^2$ whereas  its perimeter is
given by $P(r)=2\pi \sinh(\kappa   r)/\kappa$, so that for   large $r$,
both $A(r)$ and $P(r)$ grow as $\exp(\kappa r)$.

As a  consequence  of the above  property,  boundary effects are never
negligible, even in the thermodynamic  limit. Generally speaking,  the
statistical  mechanics of fluids  on $H^2$  involves multiple integral
over space, as in Eq. (\ref{eq:GC}).  Integration is performed on both
the  ``bulk''  and the ``boundary region''  of  the system, the latter
being taken as a region of finite width (or a width whose ratio to the
linear  size  of the  system  goes  to  zero when  the  latter goes to
infinity)  near  the boundary in  an  otherwise very large sample. The
boundary contribution  of  course depends  on the  boundary  condition
imposed on the system, but it is never negligible compared to the bulk
contribution. This  is  already  true for   the ideal-gas  limit.  The
canonical  partition function $Z_1$  is proportional to the total area
$A$; however, the proportionality constant is not a pure bulk property
and depends on the type of boundary condition\cite{modes:041125}.

In this work, we are only  interested in the \textit{bulk} behavior of
fluids in $H^2$,  which is what can  be directly compared to fluids in
Euclidean  space.  One way  to eliminate unwanted  boundary effects in
$H^2$        is     to       consider        periodic         boundary
conditions\cite{Sausset:2007}.  This has been   done for  instance  by
Modes and Kamien for calculating the properties of a gas of hard disks
in the low-density limit  near ideal behavior\cite{modes:041125}.   We
shall take   here a simpler  route that  has already  been followed in
previous studies of spin   and field-theoretical models  in hyperbolic
geometry\cite{Rietman1992,Angl`esd'Auriac2001,Doyon2004a}:          we
implicitly  restrict      all   integrals       appearing    in    any
statistical-mechanical expression  to  the   ``bulk''  of the  sample,
\textit{i.e.}  the fluid far enough from  the  boundary. All resulting
thermodynamic   and  structural quantities  will therefore   be dubbed
``bulk''  ones. We stress  again  that the present problem  associated
with the thermodynamic  limit in hyperpolic geometry  is a general one
which  is not  related  to  the  presence  of long-range  correlations
between particles.  (This point will be addressed further down.)

From the (bulk) grand-canonical partition function in Eq. (\ref{eq:GC}), one obtains the (bulk) thermodynamic pressure as
\begin{equation}\label{eq:3}
\beta P=\frac{\partial ln\Xi}{\partial A}|_{T,\mu}.
\end{equation} 

For    short-range interactions, the    low-density  expansion of  the
equation of state can be put in the usual form\cite{modes:041125},
\begin{equation}\label{eq:4}
\frac{\beta P}{\rho}=1+\sum_{j\geq2}B_j(T)\rho^{j-1},
\end{equation}
where $\rho=N/A$  and  $B_j$ is  the  $j$th  virial  coefficient. When
restricting the  calculation  to the bulk  of  the system,  the second
virial coefficient $B_2$ can be expressed as
\begin{equation}\label{eq:5}
B_2=-\frac{1}{2}\int dA \, f(r),
\end{equation}
where $dA$ is given by Eq. (\ref{eq:2}) and $f(r)=(e^{-\beta u(r)}-1)$
is the  Mayer function. (In the  calculation  of $B_2$ for short-range
interactions, it is easy  to see that  one obtains the same result  by
restricting the   system to the  bulk  region  and  by using  periodic
boundary conditions\cite{modes:041125}.) As a result,
\begin{equation}
B_2=\pi \int_0^\infty dr \, \frac{\sinh(\kappa r)}{\kappa}f(r),
\end{equation}
which for hard disks of diameter $\sigma$ simplifies to\cite{modes:041125}
\begin{equation}\label{B_2HD}
B_2=\pi\frac{(\cosh(\kappa \sigma)-1)}{\kappa^2}.
\end{equation}
The  above expression correctly    reduces  to the Euclidean    result
$B_2=\pi\sigma^2/2$ when $\kappa\rightarrow 0$.

To go  beyond the low-density regime and  describe the liquid phase, a
natural strategy is  to relate the pressure  to the pair  distribution
function\cite{Hansen1986}. However,  the  hyperbolic  geometry  prevents a
direct derivation through the  Clausius function\cite{Hansen1986}. We have
instead     used     a       variant   of     the     Green-Bogoliubov
method\cite{widomrowlinson82}.    To  calculate    the     derivative   in
Eq. (\ref{eq:3}), we choose to perform an affine transformation of the
elementary area element,
\begin{equation}\label{eq:6}
dA'=(1+\xi)dA,
\end{equation}
with $\xi$ an infinitesimal  parameter. Inserting Eq. (\ref{eq:2})  in
Eq. (\ref{eq:6})  yields  the following transformation  of  the radial
coordinate $r$ to first order in $\xi$:
\begin{equation}\label{eq:7}
r'=r+\xi\frac{(\cosh(\kappa r)-1)}{\kappa\sinh(\kappa r)}.
\end{equation}
The presence of curvature leads  to a nonlinear transformation of  the
coordinate,   contrary   to    what     occurs  in    flat   Euclidean
space\cite{kierlik:4256}. We then consider the infinitesimal variation
of $ln(\Xi)$ generated by the affine transformation,
\begin{equation}\label{eq:8}
\delta \ln(\Xi)=\xi\langle N\rangle -\frac{\beta}2 \int dA_1\int dA_2\,
\rho^{(2)}({\bf r}_1,{\bf r}_2)\delta u(r_{12}),
\end{equation}
where $\rho^{(2)}({\bf r}_1,{\bf r}_2)$ is the (bulk) two-particle density. By using Eq. (\ref{eq:7}), the infinitesimal increase of the potential is found equal to
\begin{equation}\label{eq:9}
\delta u(r)=\xi\frac{\cosh(\kappa r)-1}{\kappa\sinh(\kappa r)}u'(r)
\end{equation}
where $u'(r)$ denotes the derivative of $u(r)$ with respect to $r$.

Taking advantage of  the homogeneity and the  isotropy of ${H}^2$ (far
from the boundary), we then obtain the bulk thermodynamic pressure as
\begin{equation}\label{eq:10}
\frac{\beta P}{\rho}=1 -\frac{\pi \beta\rho}{\kappa^2}\int_0^\infty dr \,g(r)(\cosh(\kappa r)-1)
 u'(r),
\end{equation}
where we have used that $\rho^{(2)}({\bf r}_1,{\bf r}_2)=\rho^2 g(r)$,
$g(r)$ being the bulk  radial distribution function\cite{Hansen1986}. This
result generalizes the virial  equation of state obtained in Euclidean
space. The latter  is recovered when  taking the $\kappa\rightarrow 0$
limit in Eq. (\ref{eq:10}):
\begin{equation}\label{eq:11}
\frac{\beta P}{\rho}=1 -\frac{\pi \beta\rho}{2}\int_0^\infty dr \, r^2 g(r)  u'(r).
\end{equation}

At  large distance, $g(r)$  goes to $1$,   and Eq. (\ref{eq:10}) shows
that  the pressure is only defined  if $\int_R^\infty dr (\cosh(\kappa
r)-1)  u'(r)  $   is  finite,  with   $R$  some    irrelevant  cut-off
distance. This imposes that  the pair  potential $u(r)$ decays  faster
than $\exp(-\kappa   r)$ when  $r\rightarrow \infty$. This  requirement
prevents using any  interaction potential   with an algebraic   decay,
contrary to the  $d-$dimensional  Euclidean  case where all potentials    $u(r)\propto
r^{-n}$ with $n\ge d$ lead to a well-defined thermodynamic limit (for the
Coulombic interaction,  see below). The above  condition  is of course
satisfied  for  hard disks, but also  for   the one- and two-component
plasma  models because,   as   will be discussed   later,  the Coulomb
potential  decreases  exponentially fast at     large distance in  the
hyperbolic plane.

In the  case of hard  disks, after  introducing the auxiliary function
$y(r)=g(r)e^{\beta u(r)}$, we can express the pressure as
\begin{equation}
\frac{\beta P}{\rho}=1 +\frac{\pi \rho}{\kappa^2}\int_0^\infty dr (\cosh(\kappa \sigma) -1) y(r) \frac{d}{dr}\exp(-\beta u(r))
\end{equation}
and using $\frac{d}{dr}\exp( -\beta u(r))=\delta(r-\sigma)$  finally leads to
\begin{equation}\label{eq:12}
\frac{\beta P}{\rho}=1 +\frac{\pi  \rho}{\kappa^2} (\cosh(\kappa \sigma) -1)g(\sigma^{+}).
\end{equation}
At low density, the radial distribution function at contact $g(\sigma^{+})$
goes to one, so that one recovers the virial expansion in Eq. (\ref{eq:4}) to first order with the second virial coefficient given by Eq. (\ref{B_2HD}).

\section{Integral equations}\label{sec:integral-equations}

In liquid-state theory, approximate  integral equations are a standard
tool to describe the structure and the thermodynamics of a system. The
approach is based   on   the (exact) Ornstein-Zernike  equation   that
relates the radial distribution function $g(r)$, or more precisely the
so-called  pair  correlation  function  $h(r)=g(r)-1$,  to  the direct
correlation  function  $c(r)$ which   is    obtained from  the  second
functional derivative  of the  grand  potential with respect to  local
density   fluctuations\cite{Hansen1986,stell75}.  In   the  hyperbolic  plane,  the
Ornstein-Zernike equation reads
\begin{equation}\label{eq:13}
h(r)=c(r)+\rho \int dA' h(r')c(t(\mathbf{r},\mathbf{r'}))
\end{equation}
where $t(\mathbf{r},\mathbf{r'})$  is the  modulus of the displacement
associated   with   an    element  of   the    hyperbolic  translation
group\cite{Beardon1983}.          (In          Euclidean        space,
$t(\mathbf{r},\mathbf{r'})=|\mathbf{r}-\mathbf{r'}|$.)     Again,  Eq.
(\ref{eq:13}) should be considered as a bulk result, in which boundary
effects have been removed when taking the thermodynamic limit.

The Ornstein-Zernike   equation is more   conveniently expressed after
applying  a Fourier-Helgason transform  (that generalizes  the Fourier
transform of the Euclidean space):
\begin{equation}\label{eq:14}
\tilde{h}(k)=\tilde{c}(k)+\rho \tilde{c}(k)\tilde{h}(k).
\end{equation}
Note  that  for an isotropic  function  $h(r)$ as considered here, the
Fourier-Helgason    transform   becomes           a        Mehler-Fock
transform\cite{Terras:1985}. For more details, see Appendix A.

The  basis of the integral-equation approach  for  liquids is that the
direct   correlation function $c(r)$  has a   simpler structure and is
shorter-ranged than $h(r)$ and, as a consequence, is a better starting
point for  approximations. Common approximations are the Percus-Yevick
(PY)  and  hypernetted   chain (HNC)  closures in  which   the  direct
correlation function is taken as\cite{Hansen1986}
\begin{eqnarray}\label{eq:15}
c(r)=(1+\gamma(r))(\exp(-\beta u(r)) -1) & {\rm (PY)}\\\label{eq:16}
c(r)=\exp(-\beta u(r) +\gamma(r))-(1+\gamma(r)) & {\rm (HNC)},
\end{eqnarray}
where $\gamma(r)=h(r)-c(r)$.

In Euclidean geometry and for hard spheres, the PY closure (the direct
correlation function is then zero  beyond $\sigma$) is amenable to  an
analytical         solution     in       spaces        of          odd
dimensions\cite{Hansen1986,Leutheusser1984,robles:016101,rohrmann:051202,robles:219903,bishop:034506}
and compares rather well with computer simulation results. Recently, a
semi-analytical solution  of the  PY  integral equation has  also been
obtained                in           two-dimensional         Euclidean
space\cite{adda-bedia:184508,adda-bedia:144506}: the first ten  virial
coefficients have    been calculated  and  compared  to   Monte  Carlo
results. In addition, the   PY  approximation for hard discs has    been
numerically solved on a sphere\cite{Lishchuk2006}.

Due to the approximations involved in  the integral-equation approach,
going from  the structure  of the  fluid to  its  (bulk) thermodynamic
properties depends  on the  route  chosen.  Once the  pair correlation
function is  known, one may  find  the pressure  with the  help of the
virial    equation,  Eq.    (\ref{eq:10}),   but  also    through  the
compressibility relation  or      that  for  the     excess   internal
energy\cite{Hansen1986}. For instance, the compressibility relation reads
\begin{equation}\label{eq:compress}
\rho k_B T\chi_T= 1+ \rho \int dA\, h(r),
\end{equation}
where again only bulk contributions are considered.

At this  point, it is  important to stress  a peculiar feature  of the
Fourier-Helgason transform:   contrary to   what occurs in   Euclidean
space,  $\int dA\,  h(r)   \neq  \tilde{h}(0)$. Indeed, by  using  the
results of Appendix A, one finds that
\begin{eqnarray}\label{eq:neq}
\nonumber \tilde{h}(0)&= \frac{2\pi}{\kappa}\int_0^\infty dr \sinh(\kappa r) P_{-1/2}(\cosh(\kappa r)) h(r) \\& \neq \frac{2\pi}{\kappa}\int_0^\infty dr \sinh(\kappa r) h(r),
\end{eqnarray}
where $ P_{-1/2}(x)$ is a Legendre function of the first kind (conical function).
As a consequence,  the bulk compressibility in $H^2$  is  not given by
the $k=0$ value of   the    Fourier-Helgason transform of the     pair
correlation  function.  On the other    hand, neglecting the  boundary
effects, one still has
\begin{equation}\label{eq:compress2}
\rho k_B T\chi_T= \frac{1}{1- \rho \int dA\, c(r)},
\end{equation}
so that the bulk pressure can be obtained by a thermodynamic integration according to
\begin{equation}\label{eq:17}
\frac{\beta P}{\rho}= 1-\frac{1}{\rho}\int_0^\rho d\rho'\rho' \int dA\, c(r;\rho')
\end{equation} 
where $c(r,\rho)$  denotes the direct correlation function at the density $\rho$.

For  the hard-sphere fluid in Euclidean space, the  Percus-Yevick  integral equation provides a very accurate equation of state when the pressure is calculated with the following empirical rule\cite{Hansen1986}:
\begin{equation}\label{eq:18}
P=\frac{2P_c+P_v}{3}
\end{equation}
where $P_c$ and $P_v$ denote the ``compressibility'' and ``virial'' pressures obtained from Eqs. (\ref{eq:17}) and (\ref{eq:11}), respectively.

Note  finally that  according  to  Eq. (\ref{eq:compress}), a   finite
compressibility implies that $\int dA\, h(r)$ is finite, which in turn
requires that $h(r)$ decays at large distance faster than $\exp(-\kappa
r)$. A gas-liquid critical  point with a diverging compressibility may
thus    occur   with a pair    correlation    function that  decreases
exponentially as  $\exp(-\kappa r)$.  Assume  for instance  that $h(r)$
behaves as $\exp(-r/\xi)$ at large distance. One then finds, with again
the restriction that the integration is  only over the bulk region far
from the boundary, that
\begin{equation}\label{eq:critical}
\int_{r<L} dA\, h(r) \sim \frac{(\kappa \xi)^2}{1-(\kappa \xi)^2}\left[1 - \left( \frac{1+\kappa \xi}{2 \kappa \xi}\right) e^{-(\frac{1}{\xi}- \kappa)L} \right] + reg,
\end{equation}
when $L \rightarrow + \infty$, where $reg$ denotes regular, finite terms.
The above expression diverges  when  $\kappa \xi \rightarrow 1^-$.  If
$\xi$ is a regular function of $T$ and $\rho$ when $\xi <
\kappa^{-1}$, this suggests that the  critical behavior found near  to
$\xi_c=   \xi(T_c,\rho_c)= \kappa^{-1}$  is  of  mean-field type, with
classical exponents. In  particular, the bulk compressibility $\chi_T$
would diverge as $1/\left(1-(\kappa  \xi)^2\right) \sim (T-T_c)^{-1}$,  implying that
the  critical exponent $\gamma =  1$. In addition, assuming that $h(r)
\sim \exp(-\kappa r)$ at large distance at the critical point and using
the   known   expression    of  the    Fourier-Helgason  transform  of
$1/\cosh(\kappa  r)$\cite{Terras:1985},  it    is easy to  derive  that  the
Fourier-Helgason  transform  $\tilde{h}(k)$  of  the  pair correlation
function is finite at    $k=0$, even if the   compressibility diverges
(which  is a dramatic  illustration of Eq.  (\ref{eq:neq})), and has a
regular expansion  in $(k/\kappa)^2$ when $(k/\kappa)  \rightarrow 0$.

The expected mean-field nature  of  the gas-liquid critical  point  in
$H^2$ has  the  very same origin  as  the mean-field character of  the
critical      point   of    the       Ising    model in     hyperbolic
geometry\cite{Rietman1992,Angl`esd'Auriac2001,Doyon2004a}:         the
divergence of  the compressibility/susceptibility is controlled by the
exponential growth  with distance of   the differential  surface  area
associated with the  hyperbolic metric. An  approximate description of
the  gas-liquid  critical behavior will  also   be provided in section
\ref{sec:trunc-lenn-jones}.

\section{Coulombic systems}\label{sec:coulombic-systems}

We now consider systems  of  particles interacting through  a  Coulomb
potential,    as     was   done       before    by  Jancovici      and
coworkers\cite{Jancovici1998,fantoni2003,jancovici2004}.      For    a
monodisperse system  (particles with equal charges  of the same sign),
it  is necessary  to  add a charged uniform  background  such that the
global electroneutrality  is satisfied,    thereby allowing  a  proper
thermodynamic limit.  In  the limit   of point-like  particles,   this
corresponds to the one-component plasma model (OCP).  When considering
a binary mixture of two species  of oppositely charged particles (with
charges $\pm q$) in  equal concentrations, it is  possible to obtain a
well-defined    thermodynamic limit   without   introducing  a charged
background: in the limit of point-particles, this is the two-component
plasma model (TCP). In this latter case, the system  is stable at high
temperature, down to a  temperature  at which  a collapse of  pairs of
opposite charges occurs. When the charged particles have an additional
hard-core interaction and at  low charge density, the system undergoes
a Kosterlitz-Thouless transition  at  a lower temperature  $T_c=q^2/(4
k_B)$\cite{PhysRevB.33.499,PhysRevB.55.522}.

The  Coulomb potential  is formally  defined   as the solution  of the
Poisson equation,  which  is related to the  Green's   function of the
following inhomogeneous Laplace equation: 
\begin{equation}\label{eq:Poisson}
\Delta v(r,\phi)= -2\pi\delta^{(2)}(r,\phi),
\end{equation}
where $\Delta$ is the Laplacian operator and $\delta^{(2)}(r,\phi)$ the Dirac distribution in the appropriate manifold. In the  Euclidean plane and for a function depending on the radial coordinate only, $\Delta$ is given by
\begin{equation}
\Delta v(r)=\frac{1}{r}\frac{\partial}{\partial r}\left( r \frac{\partial v(r)}{\partial r}\right),
\end{equation}
and the solution  of Eq. (\ref{eq:Poisson}) is
\begin{equation}
v(r)=-\ln(r/L)
\end{equation}
where $L$  is  an arbitrary length.  Therefore in  two dimensions, the
Coulombic interaction potential diverges  with  the distance, even  if
the   associated  force remains  a  (slow)  decreasing function of the
distance.

On the other hand, in the hyperbolic plane, the Laplacian acts on a function of the radial coordinate as
\begin{equation}\label{eq:19}
\Delta v(r)=\frac{1}{\sinh(\kappa r)}\frac{\partial }{\partial r}
\left(\sinh(\kappa r)\frac{\partial v(r)}{\partial r}\right).
\end{equation}
The solution of the inhomogeneous Laplace equation is then given by
\begin{equation}\label{eq:20}
v(r)=-\ln\left(\tanh\left(\frac{\kappa r}{2}\right)\right),
\end{equation}
so that the pair interaction between  two point particles with charges
$q_1$ and $q_2$ at a geodesic distance $r$ is equal  to $q_1 q_2 v(r)$
with $v(r)$  given above. The  curvature of the hyperbolic  plane thus
introduces a screening of  the Coulomb interaction. Indeed, $v(r)$ now
decreases as $\exp(-\kappa r)$ at large distance beyond $\kappa^{-1}$.

It is possible to derive  analytical expressions for the thermodynamic
quantities  in   two  cases: the    high-temperature limit, where  the
Debye-Hückel approximation is asymptotically exact, and the particular
value of the temperature $T=q^2/(2k_B)$, where a large amount of exact
results have been obtained both in the Euclidean and in the hyperbolic
plane.  In the following  we shall consider  the two models introduced
above, the OCP  and the TCP. (To extend  the investigation beyond  the
two  cases considered here,  one could  use  the HNC integral equation
which is known   to give a  good  description of  the pair correlation
function in flat  space\cite{Hansen1986},  but  this  is left  for  future
work.)

\subsection{One-component plasma (OCP)}\label{sec:one-component-plasma}
For the OCP,  the charged background  provides  a uniform neutralizing
contribution.   Assuming  a  perfect  compensation amounts to  replace
$g(r)$ by $h(r)=g(r)-1$  in the expression  of the virial equation  of
state.  After taking the derivative of Eq. (\ref{eq:20}), the equation
of state can be written as
\begin{equation}\label{eq:21}
\frac{\beta P}{\rho}=1 +\frac{\pi \beta q^2\rho}{\kappa}\int_0^\infty dr h(r)\frac{(\cosh(\kappa r)-1)}{\sinh(\kappa r)},
\end{equation}
where, we  recall, $P$ is   the bulk  thermodynamic pressure. (For   a
discussion of the different definitions of pressure  in the OCP in the
hyperbolic plane,  see Fantoni  \textit{et al.}\cite{fantoni2003}.) We
shall show that the  above equation of  state reduces to known results
in the appropriate limits.

Note that generalized Stillinger-Lovett sum rules  are  satisfied   by  the
OCP in the hyperbolic plane\cite{jancovici2004}, namely:
\begin{equation}
\int dA\, h(r) =-1,
\end{equation}
\begin{equation}
\frac{4\pi\rho\beta}{\kappa^2} \int dA \, h(r) \ln(\cosh(\kappa r/2)=-1.
\end{equation}
 These rules express the fact that the system is a conductor (in which
 therefore electro-neutrality and screening hold).

In the  high-temperature limit, the Debye-Hückel approximation becomes
asymptotically exact and   provides an analytical  expression  for the
pair correlation function $h(r)$\cite{Jancovici1998,jancovici2004}:
\begin{equation}\label{eq:22}
h(r)=-\beta q^2Q_\nu(\cosh(\kappa r)),
\end{equation}
where $q$ is the charge  of the particles and   $Q_\nu$ is a  Legendre
function of the second kind with an index $\nu$ given by
\begin{equation}\label{eq:23}
\nu=-\frac{1}{2}+\sqrt{\frac{1}{4}+2\frac{\pi\beta\rho q^2}{\kappa^2}}.
\end{equation}
One can check that the expression in Eq. (\ref{eq:22}) satisfies the two above sum rules.

We can  now combine the generalized  virial equation of  state that we
have  derived with the Debye-Hückel   expression for $h(r)$. Inserting
Eq. (\ref{eq:22}) in Eq. (\ref{eq:21}) leads to the following equation
of state:
\begin{equation}
\frac{\beta P}{\rho}=1 -\frac{\pi (\beta q^2)^2 \rho}{\kappa^2}\int_1^\infty dx \frac{Q_\nu(x)}{x+1}.
\end{equation}
For a  fixed nonzero curvature ($\kappa  >0$), Eq. (\ref{eq:23}) shows
that $\nu\rightarrow  0$     in the  high-temperature  limit   ($\beta
\rightarrow             0$);             knowing                  that
$Q_0(x)=\frac{1}{2}\ln\left(\frac{1+x}{1-x}\right)$ and using that
\begin{equation}
\int_1^\infty \frac{\frac{1}{2}\ln\left(\frac{1+x}{1-x}\right)}{1+x}=\frac{\pi^2}{12},
\end{equation}
one finally obtains that
\begin{equation}\label{OCP}
\frac{\beta P}{\rho}=1 -\frac{\pi^3(\beta q^2)^2 \rho}{12\kappa^2}.
\end{equation}
This  expression    coincides   with    the dominant term     in   the
high-temperature limit of the   virial expansion derived  by Jancovici
and Tellez\cite{Jancovici1998}.

On the  other hand, when  $\kappa\rightarrow 0$ (faster than $\beta^{1/2}$),
$\nu \sim \sqrt{2\pi  \beta \rho q^2}/\kappa  \rightarrow +\infty$ and
one has
\begin{equation}\label{eq:24}
\int_1^\infty dx\frac{Q_\nu(x)}{1+x}\sim \frac{1}{2\nu^2}.
\end{equation}
The pressure is then given by
\begin{equation}\label{eq:25}
\frac{\beta P}{\rho}=1 -\frac{\beta q^2}{4},
\end{equation}
which  corresponds      to    the    result   in     the     Euclidean
limit\cite{Torres2004}.

We next consider the specific value of the temperature $T=q^2/(2k_B)$,
usually expressed in terms  of  the coupling parameter  $\Gamma =\beta
q^2=2$.  After a mapping onto a  non-Hermitian fermionic field theory,
Hastings\cite{Hastings1998} has  obtained the exact expression  of the
pair correlation function for Coulomb systems  in the hyperbolic plane
for this value of the coupling. For the OCP, this provides
\begin{equation}\label{eq:26}
h(r)=-\frac{1}{\cosh(\kappa r/2)^{8\pi\rho/\kappa^2+2}}.
\end{equation}
After introducing the change of variable  $x=\cosh(\kappa r/2)$ and inserting Eq. (\ref{eq:26}) in Eq. (\ref{eq:21}), we obtain
\begin{equation}
\frac{\beta P}{\rho}=1 -\frac{2\pi \beta q^2\rho}{\kappa^2}\int_1^\infty dx \frac{1}{x^{3+\frac{8\pi \rho}{\kappa^2}}}.
\end{equation}
Since $\beta q^2=2$, the bulk thermodynamic pressure is finally given by
\begin{equation}\label{eq:gamma2}
\frac{\beta P}{\rho}=\frac{2\pi\rho+\kappa^2}{4\pi \rho+\kappa^2}.
\end{equation}
In the low-density limit, this gives $\beta P/\rho = 1- (2\pi/\kappa^2) \rho + O(\rho^2)$, which leads to the same second virial coefficient as found in Ref. \cite{Jancovici1998}.

When $\kappa\rightarrow   0$, one recovers  from Eq. (\ref{eq:gamma2})
the exact Euclidean limit
\begin{equation}
\frac{\beta P}{\rho}=\frac{1}{2}.
\end{equation}
Conversely  when $\kappa\rightarrow \infty$,  the pressure goes to the
ideal-gas limit. This  result seems to be  quite  general, namely: the
influence  of  the   inter-particle  interactions   vanishes   in  the
large-curvature limit.

\subsection{Two-component plasma (TCP)}\label{sec:two-component-plasma}

The TCP has been investigated in Euclidean space\cite{Hansen1983,samaj2000,Samaj2001,Samaj2002,Samaj2002b}. For
point    particles, the  plasma   is  stable  in the  high-temperature
conducting  phase, for a  coupling  parameter $\beta q^2<2$. Adding  a
small  hard-core repulsive potential allows   one  to obtain a  stable
phase at  lower temperatures, up to a  coupling  $\beta q^2=4$ where a
Kostelitz-Thouless transition between a conducting phase and a dipolar
phase  takes place. As  for the OCP, the model  is solvable for $\beta
q^2=2$ in various geometries\cite{GAUDIN1985,cornu:2444,tellez:8572}.

The bulk thermodynamic pressure of the TCP in the hyperbolic plane is expressable as
\begin{equation}\label{eq:27}
\frac{\beta P}{\rho}=1 +\frac{\pi \beta q^2\rho}{\kappa}\int_0^\infty dr 
(h_{++}(r)-h_{+-}(r))\frac{(\cosh(\kappa r)-1)}{\sinh(\kappa r)},
\end{equation}
where $h_{++}(r)$  and    $h_{--}(r)$  denote  the   pair  correlation
functions between particles of equal charges and particles of opposite
charges,  respectively.   In the  high-temperature/low-coupling limit,
the Debye-Hückel approximation provides an accurate description of the
conducting phase.   For the  TCP,  the linearized  Poisson equation is
similar  to that of the OCP.  The pair  correlation functions are then
given by
\begin{eqnarray}\label{eq:28}
h_{++}(r)=-h_{+-}(r)=-\frac{\beta q^2}{2} Q_\nu(\kappa r).
\end{eqnarray}
After inserting   Eq.(\ref{eq:28})  in  Eq.(\ref{eq:27}),   one  obtains an
equation  of  state in the   Debye-Hückel approximation which is equal to that
of the OCP in the same approximation, Eq. (\ref{OCP}).

For  $\beta   q^2=2$,  Jancovici and   Tellez\cite{Jancovici1998} have
obtained the exact  expressions of the  pair correlation  functions of
the TCP (in the limit of  very small hard-core interactions). However,
the calculation of the equation of state via Eq. (\ref{eq:27}) becomes
much more   involved  than for   the  OCP  without bringing  much  new
insight. Therefore, we have not pursued in this direction.

\section{Short-range potentials}\label{sec:short-range-potent}

\begin{figure}[t]
\centering
\resizebox{8cm}{!}{\includegraphics{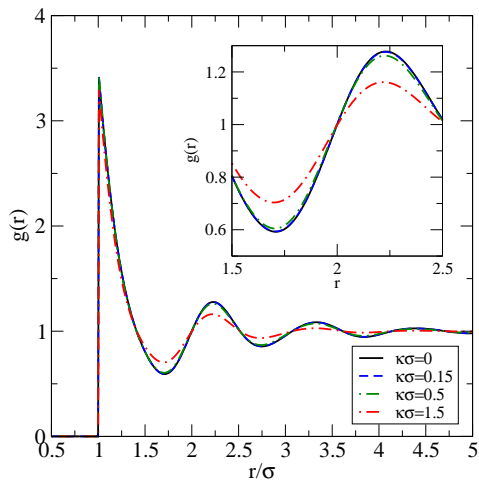}}
\caption{Radial distribution function $g(r)$ obtained from the PY equation at an area fraction $\phi =0.55$ 
for     various       values   of     the      curvature:      $\kappa
\sigma=0,0.15,0.5,1.5$. Only  for  $\kappa=1.5$,   namely a    radius of
curvature  smaller than the  particle diameter,  the structure displays
significant deviations from the Euclidean  case. The inset zooms in the
region  of  the first minimum   and   second   maximum: $g(r)$  for
$\kappa\sigma=0.15$ and  for  the Euclidean   case are  indistinguishable,
whereas   for   $\kappa\sigma=0.5$,  the   extrema  are  slightly less
pronounced.}\label{fig:1} \end{figure}

\subsection{The hard-disk fluid}\label{sec:hard-disk-fluid}

As mentioned in Introduction,  the hard-disk fluid has been considered
in the context   of geometrical frustration. Approximate  equations of
state combining the  low-density expansion and  a description of close
or        nearly          close            packings    have       been
proposed\cite{modes:235701,modes:041125,haro:116101}.

We  have  solved the  PY   integral equation, which   is known  to  be
generally  better     than    the   HNC  one   for     hard  spherical
particles\cite{Hansen1986}, for  a large density  range.  Fig.~\ref{fig:1}
shows the resulting  radial distribution function $g(r)$ for different
values of the curvature at  an area fraction $\phi=\rho\frac{2\pi (\cosh(\kappa\sigma/2)-1)}{\kappa^2}\simeq0.55$ (recall that
the   freezing  area fraction    in  the    Euclidean    plane  is    around
$0.7$\cite{santos:4622}).    We  have   also  plotted  the  Euclidean
counterpart  corresponding to $\kappa=0$.  It  is  striking that for a
range of curvature  parameter, $\kappa \sigma$  between $0$ and $0.5$,
all  curves are essentially superimposable:   in this range, the  pair
distribution function  $g(r)$  is rather insensitive to  curvature. An
evolution is on the  other hand clearly  seen for stronger curvatures,
\textit{e.g.}  $\kappa\sigma=1.5$.

\begin{figure}[t]
\centering
 \resizebox{7cm}{!}{\includegraphics{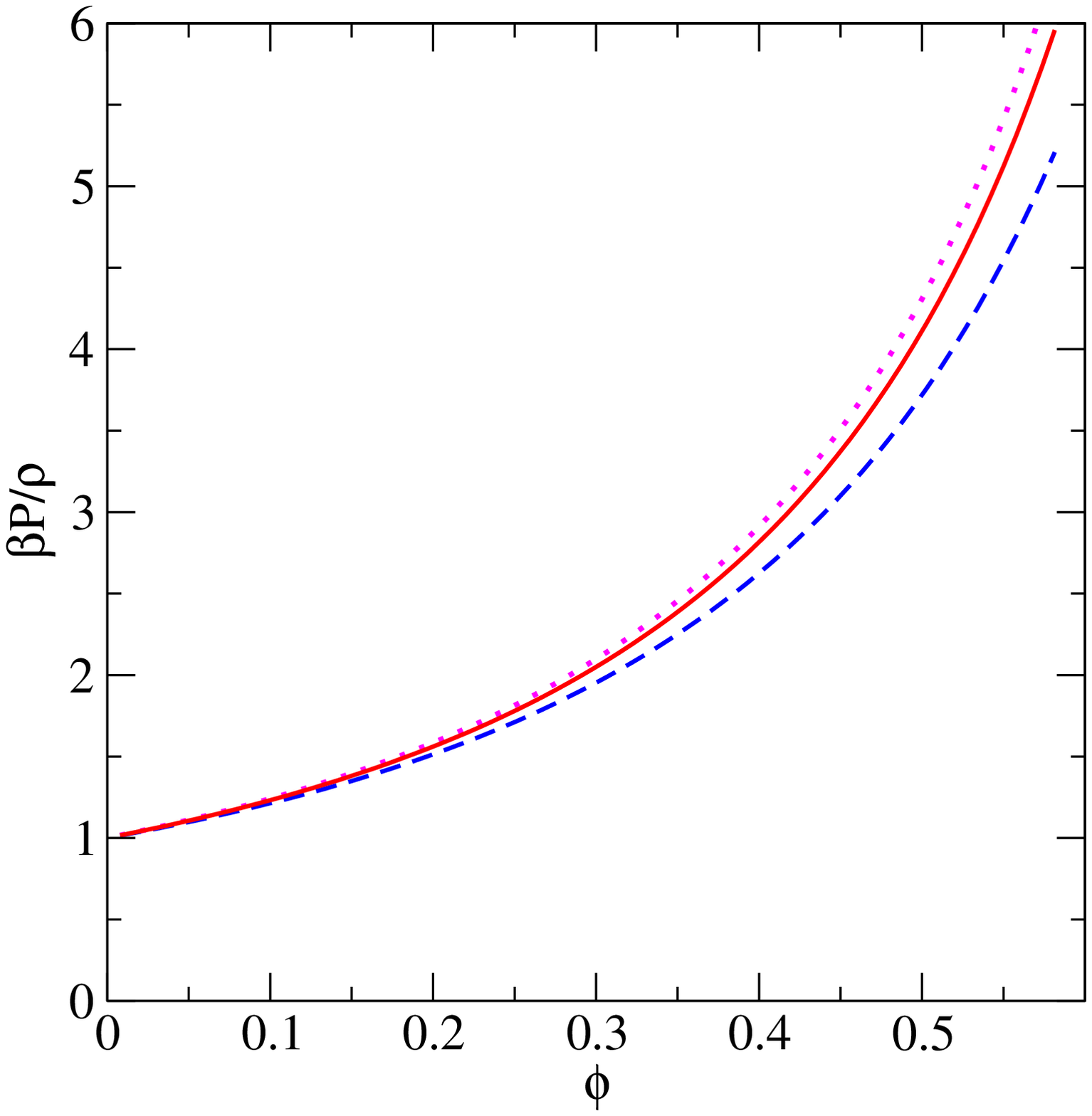}} \resizebox{7cm}{!}{\includegraphics{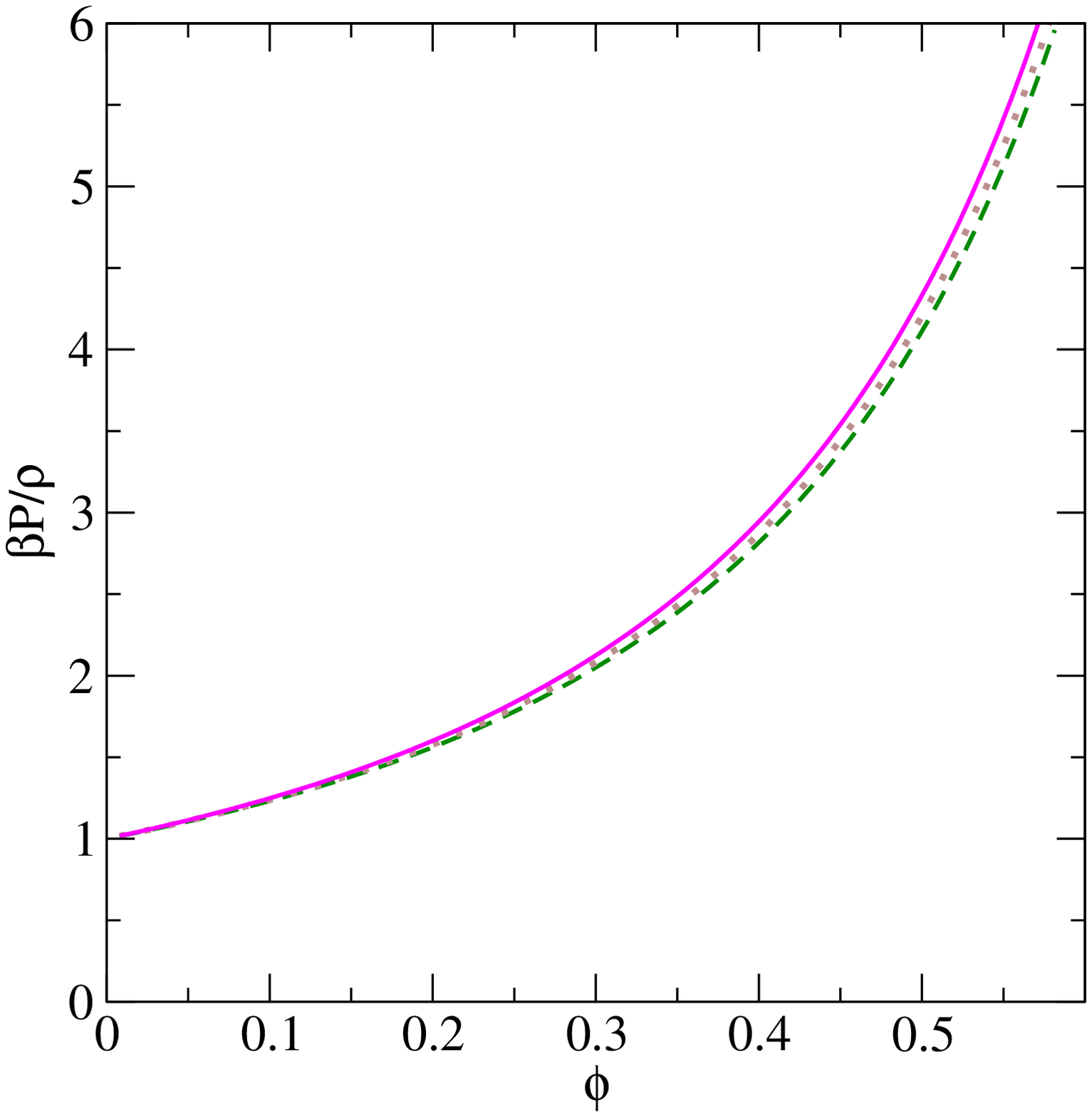}}
\caption{PY equation of state of the hard-disk fluid in $H^2$ versus area fraction $\phi$. (a) $\kappa \sigma=0.15$: 
the upper curve
 corresponds  to  the compressibility route,  the lower   curve to the
 virial route,     and the middle curve    to  the empirical   rule in
 Eq. (\ref{eq:18}). (b)  PY    equation   of state    obtained    from
 Eq.    (\ref{eq:18})        for     different  curvatures:    $\kappa
 \sigma=0.15,0.7,1.06$ from bottom to top.}\label{fig:2} \end{figure}

\begin{figure}[t]
\centering
 \resizebox{8cm}{!}{\includegraphics{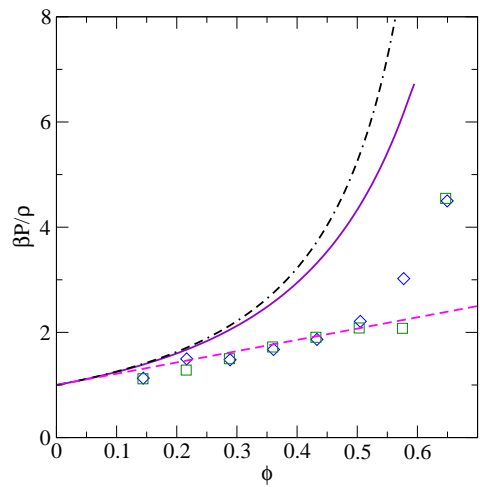}}
\caption{Equation of state of the hard-disk fluid in $H^2$ versus area fraction $\phi$:   comparison between the PY result combined with Eq. (\ref{eq:18}) (for $\kappa \sigma=1.06$) and the simulation results of Ref.~\cite{modes:041125} for $\kappa \sigma=1.060$ (green squares) and $\kappa \sigma=1.062$ (blue diamonds). The dash-dotted line is the prediction of Ref.~\cite{haro:116101} and the dashed line corresponds to the equation of state truncated at the second virial coefficient.}\label{fig:3}.
 \end{figure}

From the PY correlation  function, we have  calculated the equation of
state     through    Eqs.   (\ref{eq:11}),    (\ref{eq:17}),       and
(\ref{eq:18}).        The  results        are       illustrated     in
Fig.~\ref{fig:2}. Similarly to the Euclidean case, the compressibility
route gives a  higher   bulk thermodynamic pressure than    the virial
route, and the linear combination  given in Eq. (\ref{eq:18}) lies  in
between.    Comparison   with    the    molecular   dynamics data   of
Ref.~\cite{modes:235701,modes:041125}  and with the equation of state proposed in Ref.\cite{haro:116101} 
   is        performed     in
Fig. \ref{fig:3}. At  low area fractions simulation data and predictions share the same behavior controlled by the second virial coefficient. However,
the  situation deteriorates for even moderate area fractions.  (Note that  the number of  atoms in the simulation
cell with periodic boundary condition is very  small, always less than
$10$.)

It  is  interesting to  investigate  the local order  of the hard-disk
fluid. The geometric  frustration induced  by the  negative  curvature
describes the  impossibility of extending the  local  order present in
the dense fluid to  tile the whole  space. This is  true for a nonzero
but small enough frustration for which the local order remains that of
the Euclidean plane, \textit{i.e.} the hexagonal order. As stressed by
Rubinstein and  Nelson\cite{PhysRevB.28.6377},  there is   a
series of increasing   curvature parameters $\kappa \sigma$  for which
heptagonal, octagonal, etc..., local   order  can tile space   without
frustration.  Two points   however  are worth  stressing. First,  such
regular tilings are  only a few  of all  possible tesselations of  the
hyperbolic plane. Modes  and    Kamien\cite{modes:235701,modes:041125}
have   rather  focused   on  so-called   isostatic  regular  packings,
\textit{i.e.} tilings  of $H^2$ in which each   atom has $4$ neighbors
(the $\left\lbrace 4,5\right\rbrace$  tesselation for  instance can be
realized as a  close packing  of  disks for  a curvature  intermediate
between  the  two curvatures  studied   in their simulations:  $\kappa
\sigma=1.0612$). However, such isostatic packings have a small density
compared  to the  heptagonal $\left\lbrace7,3\right\rbrace$, octagonal
$\left\lbrace8,3\right\rbrace$,  etc...,   tilings    considered    in
Ref.~\cite{PhysRevB.28.6377}. Second, the local order in the
equilibrium  hard-disk fluid has not   actually been investigated as a
function of curvature.

\begin{figure}[t]
\centering
 \resizebox{8cm}{!}{\includegraphics{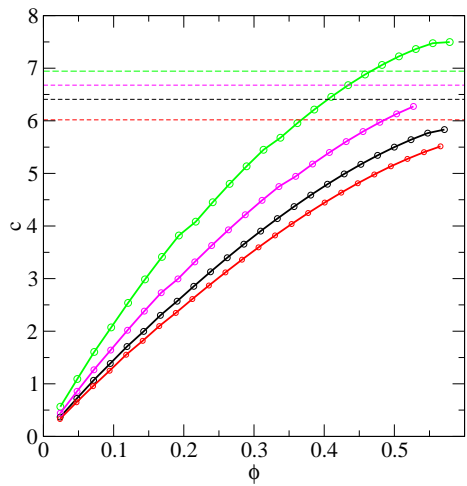}}
\caption{Average coordination number versus area fraction $\phi$  for different values of the curvature, as obtained from the PY equation. From bottom to top: $\kappa \sigma=0.15,0.7,0.9,1.06$. The dashed horizontal lines indicate the maximum values predicted by Rubinstein and Nelson\cite{PhysRevB.28.6377}.}\label{fig:4}
 \end{figure}

Our study sheds some light on  the latter question. From the knowledge
of the radial distribution  function,  we have calculated  the average
coordination number $\bar{Z}$ via the formula
\begin{equation}
\bar{Z} =2\pi\rho \int_\sigma^{\sigma_m} dr\frac{ \sinh(\kappa r)}{\kappa} g(r),
\end{equation}
where $\sigma_m$  corresponds to  the distance between  particles when
$g(r)$ attains the first minimum  beyond the first peak.  The  results
are plotted as  a  function of area fraction  for  different values   of the
curvature in  Fig.~\ref{fig:4}. As  expected, the average coordination
number increases with density and seems  to saturate (we cannot access
close packing  values with the  approximate PY integral  equation). It
also  increases  with  the  curvature  parameter.  The values at  high
density are in semi-quantitative agreement    with the prediction   of
Rubinstein and  Nelson\cite{PhysRevB.28.6377}  based   on  a
fictive ``ideal'' random close packing,
\begin{equation}
\bar{Z}(\kappa\sigma)=\frac{\pi}{\arcsin\left(\frac{1}{2\cosh\left(\frac{\kappa\sigma}{2}\right)}\right)},
\end{equation}
a  prediction that   interpolates between  the   exact values of   the
tesselations:  $\bar{Z}=6$ for   the $\left\lbrace 6,3  \right\rbrace$
tiling  when    $\kappa=0$,    $\bar{Z}=7$   for   the   $\left\lbrace
7,3\right\rbrace$ tiling  when  $\kappa \sigma=1.0905..$,  $\bar{Z}=8$
for    the      $\left\lbrace8,3\right\rbrace$      tiling        when
$\kappa\sigma=1.5285..$,  etc.... The  coordination  number is clearly
overestimated for large curvatures, \textit{e.g.} $\kappa\sigma=1.06$,
in  the PY approximation   and  we suspect  that the  validity  of the
approximate  closures  quite  generally    gets  worse as    curvature
increases.

\subsection{The truncated Lennard-Jones potential}\label{sec:trunc-lenn-jones}

The Lennard-Jones pair potential,
\begin{equation}\label{eq:LJ}
u(r)=4 \epsilon\left[ \left(\frac{\sigma}{r} \right)^{12} - \left(\frac{\sigma}{r} \right)^{6} \right]
\end{equation}
is commonly   used to model liquids  in  Euclidean space.  However, as
shown in section  \ref{sec:integral-equations}, its  power-law  nature
makes  it inappropriate for   hyperbolic geometry as no  thermodynamic
limit would   be reachable.  Physically,  the $(\sigma   /r)^{6}$ term
arises from  London dispersion  forces  and its  electrostatic  origin
suggests  that the  dependence   on the geodesic  distance  should  be
modified in the hyperbolic plane.  Indeed,  even the Coulomb potential
that  (logarithmically)   increases with  distance in  two-dimensional
Euclidean space  is  found to  decay  exponentially in the  hyperbolic
plane  with an  intrinsic screening  length which is   provided by the
radius of curvature $\kappa^{-1}$. As for the $(\sigma /r)^{12}$ term,
it is  merely a convenient way to  mimic a steeply repulsive potential
between atoms and its algebraic character has no physical foundation.

To preserve  a proper  thermodynamic   limit in $H^2$,  the   simplest
procedure    is     to   truncate  the     Lennard-Jones    potential,
Eq. (\ref{eq:LJ}), beyond some cut-off distance $r_c$: this should not
significantly alter the behavior of the liquid, provided $r_c$ is less
than the radius of curvature $\kappa^{-1}$.  This cut-off procedure is
anyhow most  often employed in  the Euclidean  case in  order to  save
computer time,  and this  will facilitate  the comparison between  the
results in  Euclidean and hyperbolic  planes. In the  Euclidean plane,
Glandt\cite{glandt:4503,glandt:4546} has  obtained     the   numerical
solution of  the PY and  HNC integral  equations for the Lennard-Jones
interaction potential.  In   this  case,  the  HNC  equation   becomes
significantly more accurate than   the  PY  equation as the    density
increases.    (A comparison with  simulations   has been performed  in
Ref.\cite{singh:5463}.)

In  the present work,  we have numerically  solved both the PY and HNC
integral equations for a  large range of  temperature and density.  As
in the case of the hard-disk fluid, we observe a very small evolution of the structure with the curvature and the differences between the PY and
HNC  closures are very small too up $\kappa\sigma\simeq 0.5$ (curves not worth showing here).

We first focus on the region  of density and  temperature close to the
gas-liquid critical   point.  A   simple iterative procedure   (Picard
method) is then inefficient for converging towards the solution of the
integral equation. It is necessary to proceed  with a more sophiscated
iterative   scheme. We have combined  the  Newton   method with a  conjugate
gradient procedure\cite{Zerah1985,belloni:8080}    (which  avoids  the
operator     inversion  otherwise      present     in    the    Newton
method\cite{Gillan1979}). We have adapted this  method to hyperbolic geometry. Details are given in Appendix B.

\begin{figure}[t]
\centering
 \resizebox{8cm}{!}{\includegraphics{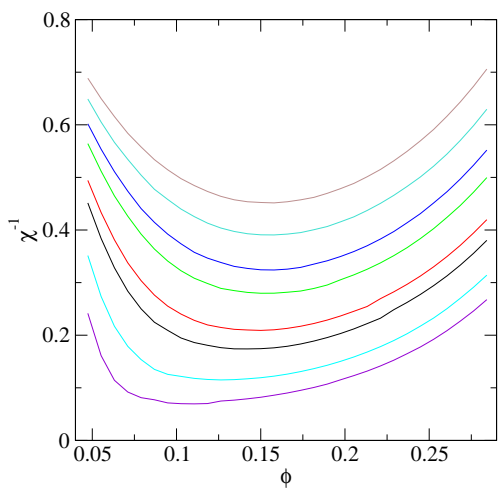}}
\caption{Inverse of the isothermal compressibility versus area fraction $\phi$ for 
the  truncated  Lennard-Jones fluid in  the PY  approximation near but
above the critical temperature: from top to  bottom , $T=0.640, 0.610,
0.580, 0.560, 0.530, 0.515, 0.490, 0.472$. The curvarture parameter is $\kappa\sigma=0.5$. }\label{fig:6} \end{figure}

\begin{figure}[t]
\centering
 \resizebox{7cm}{!}{\includegraphics{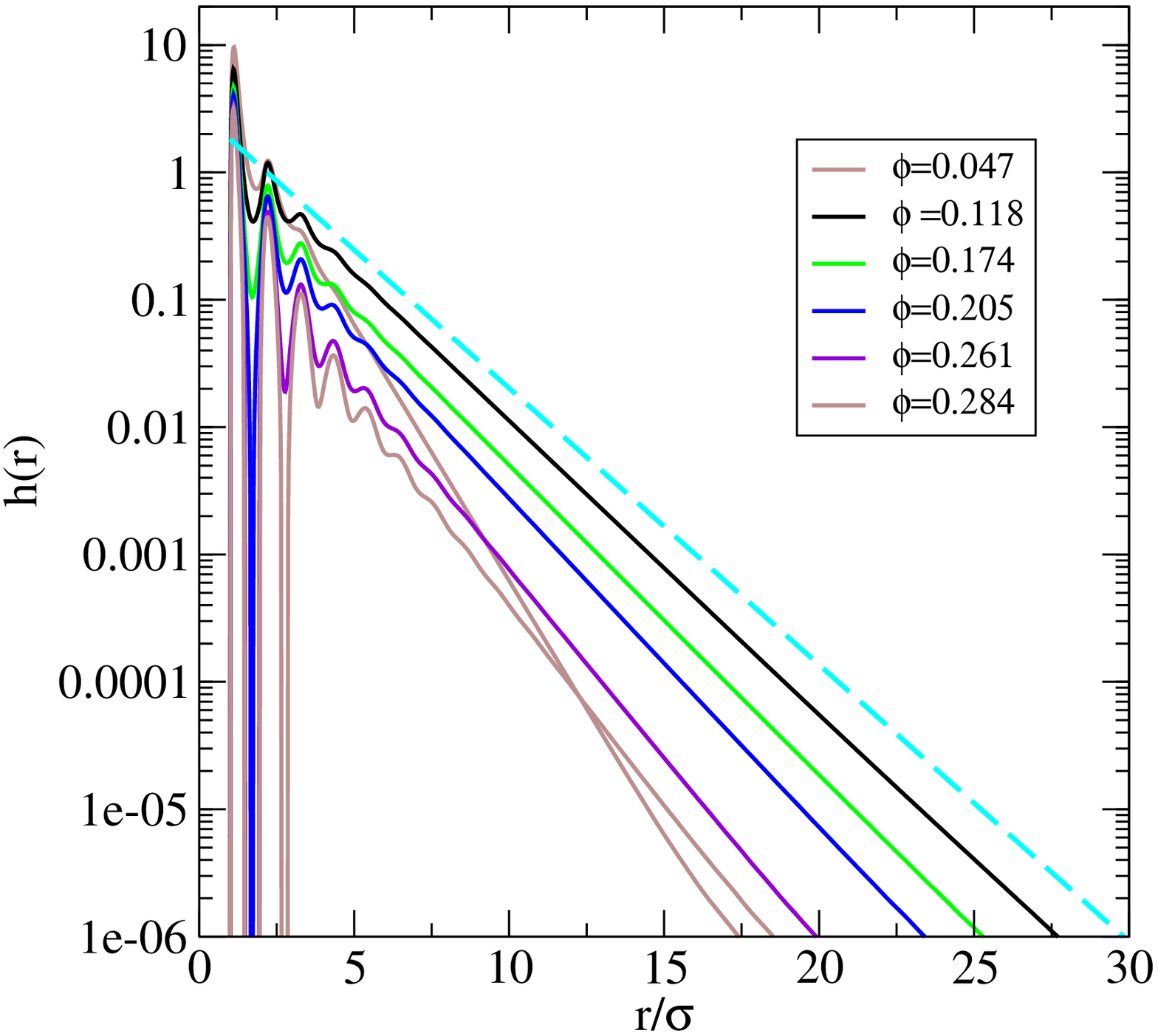}}
\resizebox{7cm}{!}{\includegraphics{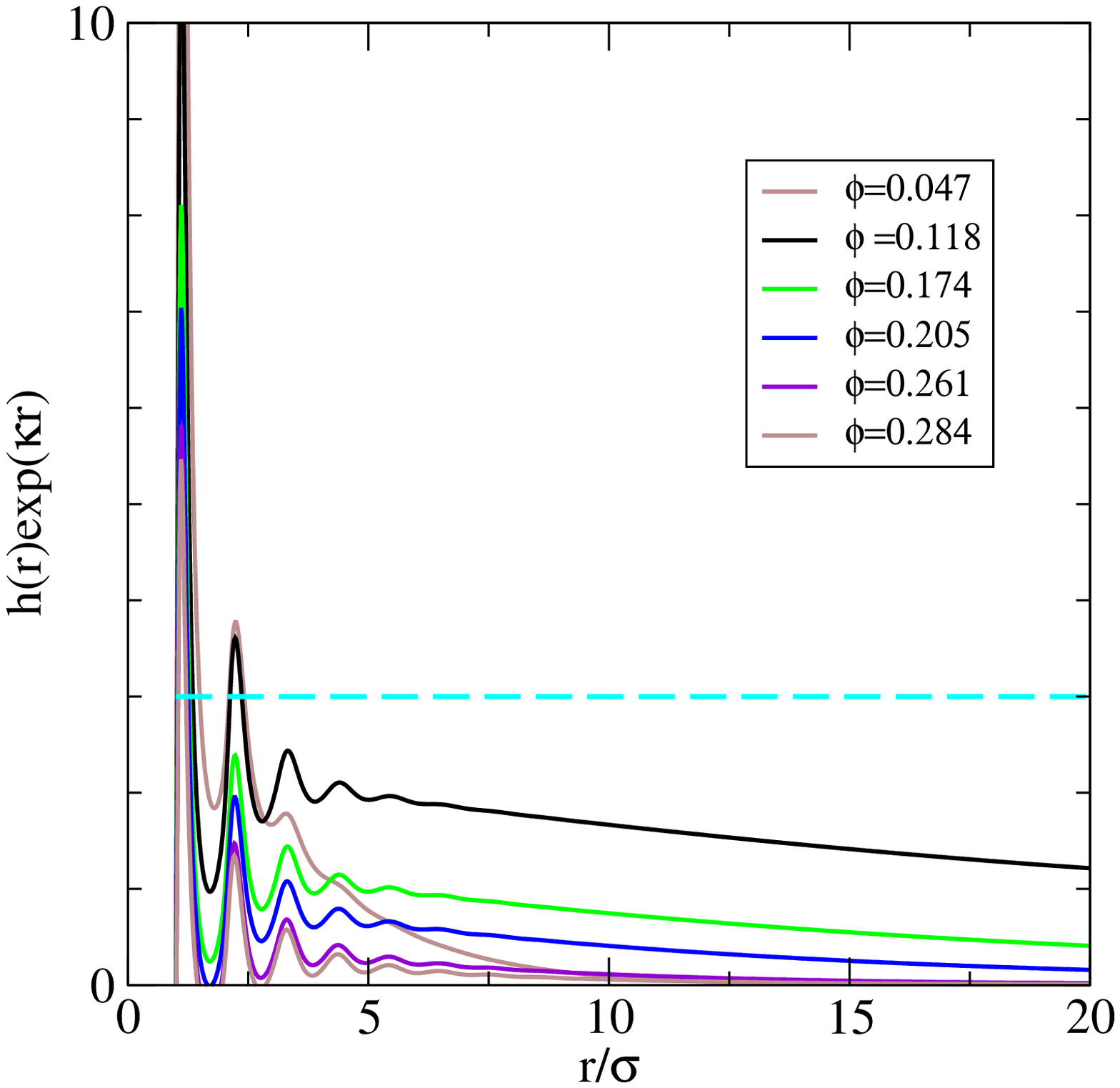}}
\caption{(a) Log-linear plot of pair correlation function $h(r)=g(r)-1$ in the PY approximation for $T=0.472$.
 The  behavior as  a function  of  $\phi$ is   non monotonous and  the
 slowest decay is for $\phi=0.118$, which  corresponds  to the maximum of
 the compressibility. The dotted line  is $\exp(-\kappa  r)$. (b) Same   data
 multiplied by $\exp(\kappa   r)$: a  convergence  towards a  constant
 plateau  at  large $r$  is  clearly  visible  as  one  approaches the
 critical point. The curvature parameter is $\kappa\sigma=0.5$.}\label{fig:7} \end{figure}

We illustrate these results for the PY closure and a reduced curvature
$\kappa\sigma=0.5$.   Differents   isotherms above   the      critical
temperature  are shown   in  Fig.\ref{fig:6}.   The   inverse  of  the
susceptibility  reaches a    minimum  at each   temperature  and   the
successive  minima approach zero  as one gets  closer  to the critical
point.  Figure~\ref{fig:7}a displays  the pair correlation functions
at   a  reduced temperature   $T=0.515$   for  several area fractions;  the
log-linear plot clearly indicates that the decay is always exponential
and faster than  $\exp(-\kappa r)$   (dotted  straight line). A   more
stringent test  of the long-distance  behavior of the pair correlation
function $h(r)$ is shown  in Fig.~\ref{fig:7}b , where  $\exp(\kappa r)
h(r)$ is  plotted as a  function of the distance  $r$. One can see the
approach to a   plateau for large   enough $r$ as one  approaches  the
critical point. (The slowest decay corresponds to the area fraction at which
the compressibility is maximum.)

\begin{figure}[t]
\centering
 \resizebox{7cm}{!}{\includegraphics{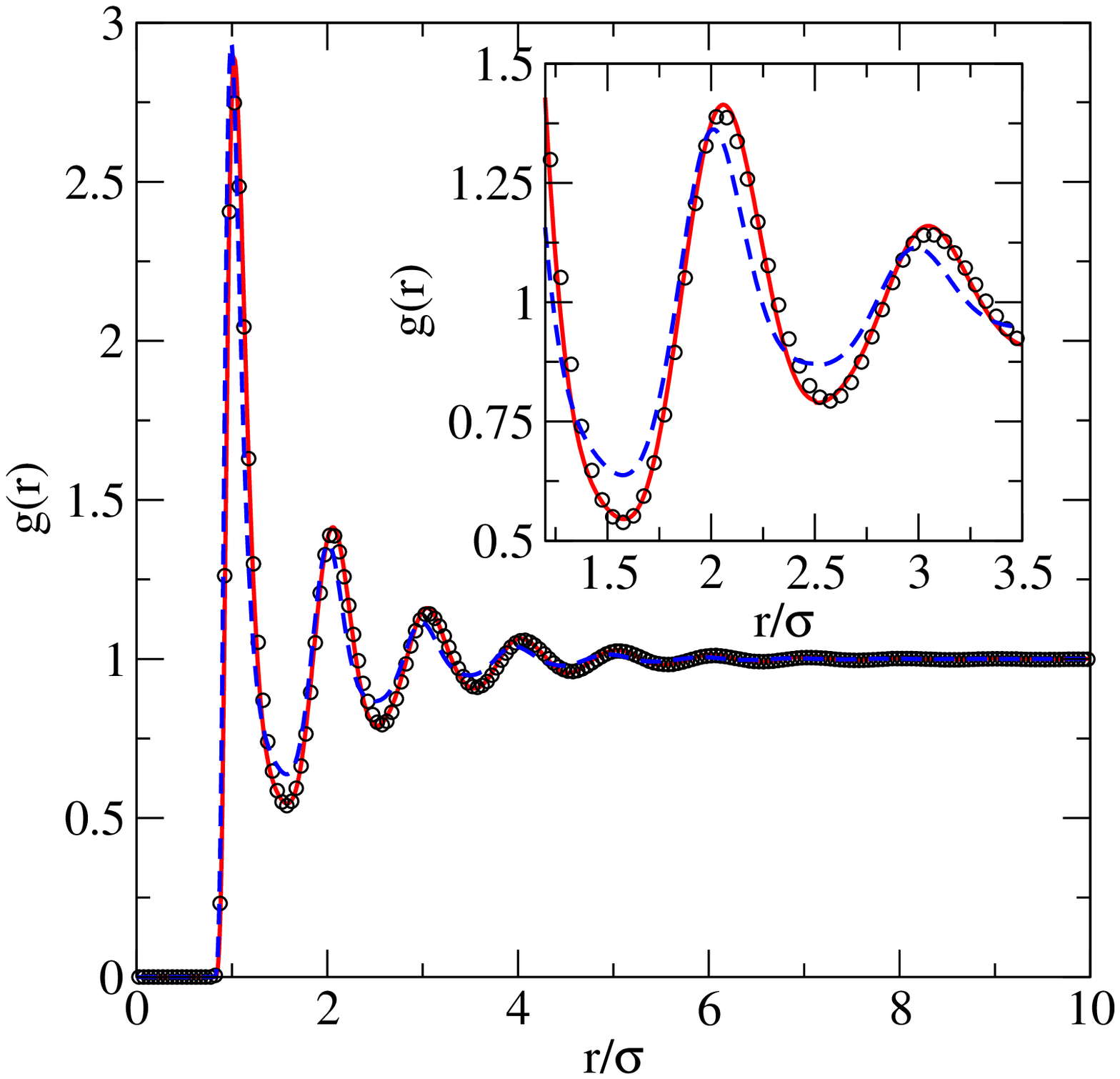}}
 \resizebox{7cm}{!}{\includegraphics{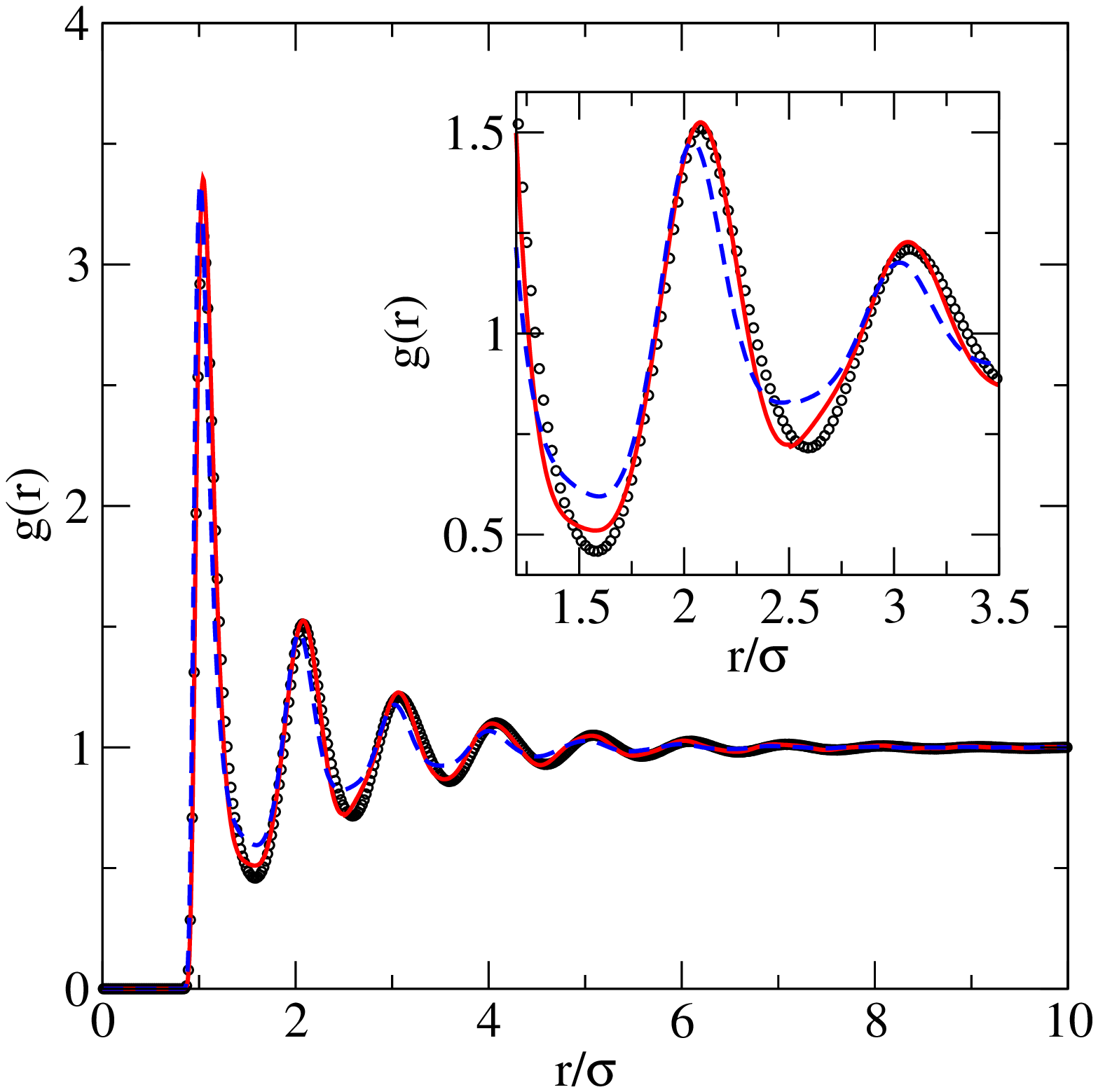}}
\caption{Radial distribution function $g(r)$ of the (truncated) Lennard-Jones
 liquid on $H^2$  for a  curvature  parameter $\kappa \sigma=0.1$, an area fraction $\phi=0.669$,
 and two different temperatures:  left,
 $T=3.259$; right,  $T=1.885$. The circles  correspond to the simulation
 data, the full (red) curve to the PY equation, and the dashed (blue)
 curve to the HNC equation. The inset zooms in  on the region near the
 first minimum and the following extrema.}\label{fig:8} \end{figure}

We have also considered the liquid at a rather high density. Recently,
we  have  performed extensive   molecular dynamics simulations  of the
monodisperse truncated Lennard-Jones liquid   in the hyperbolic  plane
(with   periodic boundary  conditions)\cite{sausset:155701}. We   have
shown that a small curvature  indeed prevents long-range hexagonal (or
hexatic) ordering and  leads  to  glass  formation as temperature   is
lowered. Most    results   were   obtained  for    a   density   $\rho
\sigma^2=0.852$.  Fig.~\ref{fig:8}   displays the  radial distribution
function $g(r)$  at an area fraction $\phi=0.669$     and a high   temperature
$T=3.259$ for a curvature parameter $\kappa \sigma=0.1$ ($\rho
\sigma^2$ is then equal to $0.852$). At this state
point,  the PY solution provides a  slightly better description of the
simulation data than the HNC equation: the HNC equation underestimates
the structure  of  the liquid at  short  distances (see  the  inset of
Fig.~\ref{fig:8}).   When  the  temperature   is lowered  to $T=1.885$
(\textit{i.e.}, still above the freezing  temperature in the Euclidean
plane      which     is      around   $T^*\simeq  0.75$     at  this
density\cite{sausset:155701}), the  HNC  and  PY closures  are  now of
comparable agreement with the simulation data, but  the maximum of the
first peak is overestimated. Overall,  the agreement appears good with
either the PY or the HNC prediction.

 \section{Conclusion}\label{sec:conclusion}

We have  formulated a statistical-mechanical  treatment for describing
the thermodynamics   and  the structure  of  fluids embedded    in the
hyperbolic plane, focusing on  the bulk behavior.  We have applied the
formalism to   the  one- and the    two-component  plasma models,  the
hard-disk  fluid  and the (truncated)   Lennard-Jones liquid  model. The
negative curvature of space provides  an intrinsic length that screens
the long-distance behavior of the  pair correlations: this is observed
both for the   Coulombic interactions  and   for the vicinity of   the
gas-liquid   critical point.  The formalism   correctly reproduces the
known results in the appropriate limits  and, via approximate integral
equations and exact thermodynamic  relations, it leads  to predictions
that compare well with existing simulation data.

A major motivation for studying  liquids in the hyperbolic plane comes
from   the theoretical description of   jamming   phenomena and  glass
formation  in   terms of   geometric  frustration.  Frustrated   local
hexagonal order in negatively curved space provides  a simple model to
study the  glass  transition.  Among the   many theories  proposed for
explaining  the glass phenomenology, several  attribute the slowing of
relaxation with decreasing temperature to the presence of a very large
number, actually exponential in  the system size, of metastable states
of  low      energy     or     free energy\cite{Goldstein1969,Stillinger1995,Kirkpatrick1989}. Techniques borrowed from  spin-glass  theory such as  the
replica  formalism   have  been applied  to    liquid models in  three
dimensions  to  check  for  the existence   of  such  a   multitude of
metastable  states, and   evidence     has  been   provided     within
mean-field-like                                           liquid-state
approximations\cite{PhysRevLett.54.1059,M'ezard1999,pz08}.   Since in some sense,   the
hyperbolic  metric induces  in statistical  systems  some  features of
mean-field models, one may wonder whether a glassforming liquid in the
hyperbolic  plane such  as    the truncated Lennard-Jones   model also
possesses  a  complex free-energy  landscape at  low  temperature. The
integral-equation  approach developed in the  present formalism can be
combined with the replica  method to address this  point. Work in this
direction is in progress.

P.V. acknowledges B. Jancovici for useful suggestions on Coulombic systems.

\appendix
\section{Harmonic analysis in the hyperbolic plane.}\label{sec:harm-analys-hyperb}

In $H^2$ it  is  possible to   define a  generalization  of the  usual
spatial  Fourier  transform,  the  Fourier-Helgason  transform.  For a
generic function $f$ of the polar  coordinates $r,\phi$, it is defined
as \cite{Terras:1985}
\begin{equation}\label{eq:29}
\tilde{f}(a,t)=\int drd\phi \sinh(r) e^{ia\phi}P^a_{-1/2+it}(\cosh(r))f(r,\phi),
\end{equation}
where $P^a_{-1/2+it}(x)$ is a Legendre function of the first kind (conical function).
The inverse Fourier-Helgason transform is given by
\begin{eqnarray}\label{eq:30}
f(r,\phi)=\frac{1}{2\pi}\sum_{a\in Z}(-1)^a\int_{t>0}&dt \, t \tanh(\pi t) e^{-ia\phi}
\nonumber\\&P^{-a}_{-1/2+it}(\cosh(r))\tilde{f}(a,t).
\end{eqnarray}

For isotropic  ($SO(2)$ invariant)  functions, the  Fourier-Helgason
transform  reduces to a Mehler-Fock transform\cite{Terras:1985}, the dependence on $a$ disappears and for a
curvature parameter $\kappa$ one obtains
\begin{equation}\label{eq:31}
\tilde{f}(k)=\frac{2\pi}{\kappa}\int_0^\infty dr \sinh(\kappa r) P_{-1/2+i\frac{k}{\kappa}}(\cosh(\kappa r)) f(r).
\end{equation}
The inverse transform is then equal to
\begin{eqnarray}\label{eq:32}
f(r)&=\frac{1}{4\pi}\int  dk \, k \tanh(\pi \frac{k}{\kappa})P_{-1/2+i
\frac{k}{\kappa}}(\cosh(\kappa r))\tilde{f}(k).
\end{eqnarray}
Let us show how one can recover the Euclidean limit when $\kappa\rightarrow 0$. Starting from the integral representation of the Legendre functions of the first kind\cite{Terras:1985},
\begin{equation}\label{eq:33}
P^\mu_\lambda(z)=\frac{\Gamma(\mu+\lambda+1)}{2\pi\Gamma(\lambda+1)}\int_0^{2\pi}du [z+\sqrt{z^2-1}\cos(u)]^\lambda e^{i\mu u},
\end{equation}
one obtains that 
\begin{eqnarray}\label{eq:34}
&P_{-1/2+i\frac{k}{\kappa}}(\cosh(\kappa r))=\nonumber\\
&\frac{1}{2\pi}\int_0^{2\pi}du [\cosh(\kappa r)+\sinh(\kappa r)\cos(u)]^{-1/2+i\frac{k}{\kappa}},
\end{eqnarray}

and when $\kappa$ goes to zero, one finds that
\begin{equation}\label{eq:35}
\frac{1}{2\pi}\int_0^{2\pi}du e^{ikr\cos(u)}=J_0(kr)
\end{equation}
where $J_0(r)$ is a Bessel function of the first kind.
Finally, the Fourier-Helgason transform, Eq.(\ref{eq:31}), becomes 
\begin{equation}\label{eq:36}
\tilde{f}(k)=2\pi\int_0^\infty rdr J_0(kr) f(r),
\end{equation}
which corresponds to the standard Fourier transform in two dimensions.

In a similar way, the inverse Helgason transform goes to
\begin{equation}\label{eq:37}
f(r)=\frac{1}{2\pi}\int_0^\infty dk J_0(kr) \tilde{f}(k).
\end{equation}

The Fourier-Helgason transform satisfies the convolution theorem: if $f$ and $g$ are $SO(2)$ invariant functions, one has 
\begin{equation}\label{eq:38}
\widetilde{(f*g)}(k)=\tilde{f}(k)\tilde{g}(k),
\end{equation}
where the star denotes the convolution product\cite{gonzalez97}.

Finally, by using  an exact Mehler-Fock  transform, we  illustrate the
result  given   in   Eq.(\ref{eq:neq}):    the  Fourier-Helgason
transform of a given function $f$ when $k\rightarrow 0$ is not equal to
the integral of this function in the hyperbolic plane. We consider
$f(r)=\cosh(\kappa   r)^{-a}$.  With the  change   of
variable  $w=\cosh(\kappa r)$, the integral  of $f$ is obviously equal
to
\begin{equation}\label{eq:39}
\int_1^\infty dw w ^{-a}=\frac{1}{1+a}
\end{equation}
whereas  the Fourier-Helgason  transform for $k=0$ is given by
\begin{equation}\label{eq:40}
\int_1^\infty w ^a P_{1/2+it}dw={\frac {\Gamma  \left( \frac{a}{2}+\frac{it}{2}-\frac{1}{4} \right) \Gamma  \left(
\frac{a}{2}-\frac{it}{2}-\frac{1}{4} \right) {2}^{a-2}}{\Gamma  \left( a \right) \sqrt {\pi }
}}
\end{equation}
where $t=k/\kappa$.
The second-order expansion of  Eq.(\ref{eq:40})  is equal to
\begin{eqnarray}
 {\frac { \left( \Gamma  \left(\frac{a}{2}-\frac{1}{4} \right)  \right) ^{2}{
2}^{a}}{4\Gamma  \left( a \right) \sqrt {\pi }}}
\nonumber\\
\fl -{\frac { \left( 
4\,\Psi \left( 1,\frac{a}{2}+\frac{3}{4} \right) {a}^{2}-4\,\Psi \left( 1,\frac{a}{2}+\frac{3}{4}
 \right) a+\Psi \left( 1,\frac{a}{2}+\frac{3}{4} \right) +16 \right) {2}^{a}
 \left( \Gamma  \left( \frac{a}{2}-\frac{1}{4} \right)  \right) ^{2}}{16 \left( 2\,a-
1 \right) ^{2}\sqrt {\pi }\Gamma  \left( a \right) }}{t}^{2}
\nonumber\\
+O \left( 
{t}^{3} \right) 
\end{eqnarray}
Apart from the limiting case $a\rightarrow  \infty$, the integral over
the whole  space is different  from the Fourier Helgason  transform at
$k=0$.  In addition, the function $f(r)$ has  a definite integral only
when $a>1$, whereas the Fourier-Helgason transform at $k=0$ is defined
for $a>1/2$. We also note that the coefficient of  the second order in
the expansion in $t$ is always negative for $a>1/2$.

\section{Solving integral equations in $H^2$: the numerical procedure.}\label{sec:solv-integr-equat}

For  solving integral  equations, the  basic   method is an  iterative
procedure, known as the Picard method, which can  be easily adapted to
systems in the  hyperbolic plane as  follows: Starting from an initial
value of $\gamma_0(r) =h_0(r) - c_0(r)$, $c_0(r)$ is obtained by using
the closure equation, Eq.(\ref{eq:15})  for PY or Eq.(\ref{eq:16}) for
HNC.   We next  perform  the Mehler-Fock  transform of $c_0(r)$.  From
$\tilde{c}_0(k)$,      and    using   Eq.(\ref{eq:14}), we   calculate
$\tilde{\gamma}_{new}(k)$     through the Ornstein-Zernike   equation,
namely,
\begin{equation}
\tilde{\gamma}(k)=\frac{\rho\tilde{c}(k)^2}{1-\rho\tilde{c}(k)}.
\end{equation}
Finally, we transform  back   $\tilde{\gamma}_{new}(k)$ to obtain    a new
function  $\gamma_1(r)$.   The solution  is taken  as  a barycenter of
$\gamma_0(r)$      and $\gamma_{new}(r)$    ,  namely  $\gamma_1(r)=\alpha
\gamma_0(r)+(1-\alpha)\gamma_{new}(r)$  where    $\alpha$    is  a  mixing
parameter. The whole procedure is repeated and we consider that convergence is reached when $|\gamma_n(r)-\gamma_{n+1}(r)|<\epsilon$ where $\epsilon\sim 10^{-8}$.

Note that  the  Mehler-Fock transform  (Fourier-Helgason transform  of
$SO(2)$ invariant functions) is  implemented by using the  Legendre or
conical functions $P_{-1/2+it}(r)$.   These functions are available in
the Gnu Scientific Library and are tabulated for saving computer time.
The number of points in
real and reciprocal space  was equal  to  $1600$, and  integration was
performed with  a Simpson rule.  In  regions of the phase diagram, far
from  the coexistence curves and   the  freezing transition line,  the
above procedure converges rapidly.

However, in  the vicinity  of  the critical  point, the  Picard method
exhibits   extremely slow convergence  and   a  Newton method is  more
appropriate for     solving  integral equations.    In  addition,  the
equations of the Newton method are not solved by a direct inversion of
the  associated operator (or  of  the large  matrix obtained after 
discretization  of the functions),  but by a more efficient procedure,
the   conjugate  gradient  methods\cite{Gillan1979,Brader2006}:   it
consists in solving  the  equations  by using an iterative method that
avoids   the inversion  of  a large   matrix (or an  operator).  The
derivation of the method is done below for  the PY closure, but it can
be easily modified for the HNC closure.

After differenciating  Eq.(\ref{eq:14})and (\ref{eq:15}),  and letting
$f(r)$ denote the Mayer function, the equations involved in the Newton
method are then given by
\begin{eqnarray}\label{eq:41}
(1-\rho \tilde{c}(k))\delta  \tilde{\gamma}(k)-\rho(\tilde{\gamma}(k)+2\tilde{c}(k))
\delta \tilde{c}(k)&=\rho(\tilde{\gamma}(k)+\tilde{c}(k))\tilde{c}(k)-\tilde{\gamma}(k)\\
\delta c(r)-f(r)\delta\gamma(r)&=-c(r)+f(r)(1+\gamma(r))\label{eq:42}
\end{eqnarray}
where $\delta c(r)(\delta \tilde{c}(k))$ and  $\delta \gamma(r)(\delta \tilde{\gamma}(k))$ are solutions of the Newton equations.

By taking the Fourier-Helgason transform of Eq.(\ref{eq:42}), one eliminates $\delta c$ and one obtains an equation of the form
\begin{equation}\label{eq:43}
A\delta\tilde{\gamma}(k)=B
\end{equation}
where 
$A$ is an operator such that
\begin{equation}\label{eq:44}
A\tilde{g}(k)=(1-\rho \tilde{c}(k))\tilde{g}(k)-\rho(\tilde{\gamma}(k)+2\tilde{c}(k))\widetilde{(fg)}(k)
\end{equation}
and  $B$ is given by
\begin{equation}\label{eq:45}
B=(-\tilde{c}(k)+\tilde{f}(k)+\widetilde{(f\gamma)}(k))
\rho(\tilde{\gamma}(k)+\tilde{c}(k))+\rho(\tilde{\gamma}(k)+\tilde{c}(k))\tilde{c}(k)-\tilde{\gamma}(k).
\end{equation}

The adjoint operator $A^\dagger$ is given by
\begin{equation}
A^\dagger\tilde{g}(k)=(1-\rho \tilde{c}(k))\tilde{g}(k)-\rho
(\widetilde{f(\gamma*g)}+2\widetilde{f(c*g)})
\end{equation}
where $*$ denotes the convolution product.

In order to solve the linear system, Eq.(\ref{eq:43}), the solution is
built on a sequence of functions  in mutually conjugate directions. We
introduce the inner product in Fourier-Helgason space:
\begin{equation}
(\tilde{X},\tilde{Y})=\int dk \frac{k}{\kappa} \tanh\left(\frac{\pi k}{\kappa}\right) \tilde{X}(k) \tilde{Y} (k).
\end{equation}

Starting with the initial conditions

\begin{eqnarray}
R_0&=&A^\dagger B\\
P_0&=&R_0
\end{eqnarray}
 the iterative procedure is given as follows:
\begin{itemize}
\item Calculate the coefficient $\alpha_k$,
\begin{equation}
\alpha_k=\frac{(R_k,R_k)}{(Ap_k,Ap_k)}.
\end{equation}
\item The residual and the solution become
\begin{equation}
R_{k+1}=R_k-\alpha_kA^\dagger A P_k
\end{equation}
\begin{equation}
X_{k+1}=X_k+\alpha_k P_k.
\end{equation}
\item The next conjugate direction is given by the relation
\begin{equation}
P_{k+1}=R_{k+1}+\beta_k P_k
\end{equation}
with
\begin{equation}
\beta_k=\frac{(R_{k+1},R_{k+1})}{(R_k,R_k)}.
\end{equation}
\end{itemize}

The iterative procedure is stopped when $(R_{k+1},R_{k+1})<\epsilon$
($\epsilon\sim     10^{-10}$ in  practice).     We  then take   the  inverse
Fourier-Helgason transform of the solution $\delta{\tilde\gamma}(k)$,
and with the help of Eq.(\ref{eq:42}), we derive $\delta c(r)$, and finally $\delta \tilde{c}(k)$ by a Fourier-Helgason transform.

We next update the  functions $\gamma$ and $c$ in both real and reciprocal
spaces :  $\gamma_1=\gamma_0+\delta  \gamma$ and  $c_1=c_0+\delta  c$.
The  inner product $b=(B,B)$ (where  $B$ is given by  Eq.(\ref{eq:45}))
is calculated by using the updated functions $\gamma$ and $c$. So, as long as
$b>\epsilon'$ with $\epsilon'\sim 10^{-8}$, the iterative procedure is repeated.

\section*{References}


\begin{thebibliography}{10}

\bibitem{Rietman1992}
R.~Rietman, B.~Nienhuis, and J~Oitmaa.
\newblock The {I}sing model on hyperlattices.
\newblock {\em J. Phys. A: Math. Gen.}, 25:6577--6592, 1992.

\bibitem{Wu1996}
C.~Chris Wu.
\newblock Ising models on hyperbolic graphs.
\newblock {\em J. Stat. Phys.}, 85:251--259, 1996.

\bibitem{Wu2000}
C.~Chris Wu.
\newblock Ising models on hyperbolic graphs {II}.
\newblock {\em J. Stat. Phys.}, 100:893--904, 2000.

\bibitem{Angl`esd'Auriac2001}
J.~C. Angl{\`e}s~d'Auriac, R.~M{\'e}lin, P.~Chandra, and B.~Dou{\c c}ot.
\newblock Spin models on non-{E}uclidean hyperlattices: Griffiths phases
  without extrinsic disorder.
\newblock {\em J. Phys. A: Math. Gen.}, 34(4):675--693, 2001.

\bibitem{Benjamini2001}
I.~Benjamini and O.~Schramm.
\newblock Percolation in the hyperbolic plane.
\newblock {\em J. Amer. Math. Soc.}, 14:487--507, 2001.

\bibitem{Shima2006}
H.~Shima and Y.~Sakaniwa.
\newblock The dynamic exponent of the {I}sing model on negatively curved
  surfaces.
\newblock {\em J. Stat. Mech.}, 2006(08):P08017, 2006.

\bibitem{Shima2006a}
H.~Shima and Y.~Sakaniwa.
\newblock Geometric effects on critical behaviours of the {I}sing model.
\newblock {\em J. Phys. A: Math. Gen.}, 39:4921--4933, 2006.

\bibitem{Ueda2007}
K.~Ueda, R.~Krcmar, A.~Gendiar, and T.~Nishino.
\newblock Corner transfer matrix renormalization group method applied to the
  {I}sing model on the hyperbolic plane.
\newblock arXiv:0704.1949, 2007.

\bibitem{Baek2007}
S.~K. Baek, P.~Minnhagen, and B.~J. Kim.
\newblock Phase transition of {XY} model in heptagonal lattice.
\newblock {\em Europhys. Lett.}, 79:26002, 2007.

\bibitem{Gendiar2008}
A.~Gendiar, R.~Krcmar, K.~Ueda, and T.~Nishino.
\newblock Phase transition of clock models on a hyperbolic lattice studied by
  corner transfer matrix renormalization group method.
\newblock {\em Phys. Rev. E}, 77:041123, 2008.

\bibitem{Baek2008b}
S.~K. Baek, H.~Shima, and B.~J. Kim.
\newblock Curvature-induced frustration in the {XY} model.
\newblock arXiv:0811.1895.

\bibitem{krcmar:061119}
R.~Krcmar, T.~Iharagi, A.~Gendiar, and T.~Nishino.
\newblock Tricritical point of the {$J_1$-$J_2$} {I}sing model on a hyperbolic
  lattice.
\newblock {\em Phys. Rev. E}, 78(6):061119, 2008.

\bibitem{baek:011124}
S.~K. Baek, P.~Minnhagen, and B.~J. Kim.
\newblock Percolation on hyperbolic lattices.
\newblock {\em Phys. Rev. E}, 79(1):011124, 2009.

\bibitem{Callan1990}
Curtis~G. Callan and Frank Wilczek.
\newblock Infrared behavior at negative curvature.
\newblock {\em Nuclear Physics B}, 340:366--386, 1990.

\bibitem{Doyon2004a}
B.~Doyon and P.~Fonseca.
\newblock Ising field theory on a pseudosphere.
\newblock {\em J. Stat. Mech.}, page P07002, 2004.

\bibitem{Belo2007}
L.~R.~A. Belo, N.~M. Oliveira-Neto, W.~A. Moura-Melo, A.~R. Pereira, and
  E.~Ercolessi.
\newblock Heisenberg model on a space with negative curvature: {T}opological
  spin textures on the pseudosphere.
\newblock {\em Phys. Lett. A}, 365:463--468, 2007.

\bibitem{jancovici2004}
B.~Jancovici and G.~T{\'e}llez.
\newblock Charge fluctuations for a {C}oulomb fluid in a disk on a
  pseudosphere.
\newblock {\em J. Stat. Phys.}, 116:205--230, 2004.

\bibitem{fantoni2003}
R.~Fantoni, B.~Jancovici, and G.~Téllez.
\newblock Pressures for a one-component plasma on a pseudosphere.
\newblock {\em J. Stat. Phys.}, 112(1):27--57, 2003.

\bibitem{Jancovici1998}
B.~Jancovici and G.~T{\'e}llez.
\newblock Two-dimensional {C}oulomb systems on a surface of constant negative
  curvature.
\newblock {\em J. Stat. Phys.}, 91:953--977, 1998.

\bibitem{modes:235701}
C.~D. Modes and R.~D. Kamien.
\newblock Hard disks on the hyperbolic plane.
\newblock {\em Phys. Rev. Lett.}, 99(23):235701, 2007.

\bibitem{modes:041125}
C.~D. Modes and R.~D. Kamien.
\newblock Geometrical frustration in two dimensions: Idealizations and
  realizations of a hard-disk fluid in negative curvature.
\newblock {\em Phys. Rev. E}, 77(4):041125, 2008.

\bibitem{haro:116101}
M.~L\'{o}pez de~Haro, A.~Santos, and S.~B. Yuste.
\newblock Simple equation of state for hard disks on the hyperbolic plane.
\newblock {\em J. Chem. Phys.}, 129(11):116101, 2008.

\bibitem{Sadoc:1999}
J.-F. Sadoc and R.~Mosseri.
\newblock {\em Geometrical frustration}.
\newblock Cambridge {U}niversity {P}ress, Cambridge, 1999.

\bibitem{Nelson:2002}
D.~R. Nelson.
\newblock {\em Defects and geometry in condensed matter physics}.
\newblock Cambridge {U}niversity {P}ress, Cambridge, 2002.

\bibitem{Nelson:1983a}
D.~R. Nelson.
\newblock Liquids and glasses in spaces of incommensurate curvature.
\newblock {\em Phys. Rev. Lett.}, 50(13):982--985, 1983.

\bibitem{PhysRevB.28.6377}
M.~Rubinstein and D.~R. Nelson.
\newblock Dense-packed arrays on surfaces of constant negative curvature.
\newblock {\em Phys. Rev. B}, 28(11):6377--6386, 1983.

\bibitem{Brito2006b}
C.~Brito and M.~Wyart.
\newblock On the rigidity of a hard-sphere glass near random close packing.
\newblock {\em Europhys. Lett.}, 76(1):149--155, 2006.

\bibitem{Kivelson:1995}
D.~Kivelson, S.A. Kivelson, X.~Zhao, Z.~Nussinov, and G.~Tarjus.
\newblock A thermodynamic theory of supercooled liquids.
\newblock {\em Physica A}, 219:27--38, 1995.

\bibitem{Tarjus:2005}
G.~Tarjus, S.~A. Kivelson, Z.~Nussinov, and P.~Viot.
\newblock The frustration-based approach of supercooled liquids and the glass
  transition: a review and critical assessment.
\newblock {\em J. Phys.: Condens. Matter}, 17(50):R1143--R1182, 2005.

\bibitem{sausset:155701}
F.~Sausset, G.~Tarjus, and P.~Viot.
\newblock Tuning the fragility of a glass-forming liquid by curving space.
\newblock {\em Phys. Rev. Lett.}, 101(15):155701, 2008.

\bibitem{Sausset:2007}
F.~Sausset and G.~Tarjus.
\newblock Periodic boundary conditions on the pseudosphere.
\newblock {\em J. Phys. A: Math. Gen.}, 40:12873--12899, 2007.

\bibitem{Hansen1986}
J.-P. Hansen and I.~R. McDonald.
\newblock {\em Theory of simple liquids}.
\newblock Academic Press, 1986.

\bibitem{widomrowlinson82}
J.~S. Rowlinson and B.~Widom.
\newblock {\em Molecular Theory of Capillarity}.
\newblock Clarendon, Oxford, 1982.

\bibitem{kierlik:4256}
E.~Kierlik, M.~L. Rosinberg, G.~Tarjus, and P.~Monson.
\newblock The pressure of a fluid confined in a disordered porous material.
\newblock {\em J. Chem. Phys.}, 103(10):4256--4260, 1995.

\bibitem{stell75}
G.~Stell.
\newblock Correlation functions and their generating functionals,.
\newblock In {\em Phase Transitions and Critical Phenomena}, volume~5b. C. Domb
  and M. S. Green (Academic, London), 1976.

\bibitem{Beardon1983}
A.~F. Beardon.
\newblock {\em The geometry of discrete groups}.
\newblock Springer-Verlag, New York, Heidelberg, Berlin, 1983.

\bibitem{Terras:1985}
A.~Terras.
\newblock {\em Harmonic analysis on symmeric spaces and applications}.
\newblock Springer-Verlag, New York, 1985.

\bibitem{Leutheusser1984}
E.~Leutheusser.
\newblock Exact solution of the {P}ercus-{Y}evick equation for a hard-core
  fluid in odd dimensions.
\newblock {\em Physica A}, 127(3):667--676, 1984.

\bibitem{robles:016101}
M.~Robles, M.~L\'{o}pez de~Haro, and A.~Santos.
\newblock {P}ercus-{Y}evick theory for the structural properties of the
  seven-dimensional hard-sphere fluid.
\newblock {\em J. Chem. Phys.}, 126(1):016101, 2007.

\bibitem{rohrmann:051202}
Ren\'{e}~D. Rohrmann and Andr\'{e}s Santos.
\newblock Structure of hard-hypersphere fluids in odd dimensions.
\newblock {\em Phys. Rev. E}, 76(5):051202, 2007.

\bibitem{robles:219903}
M.~Robles, M.~L\'{o}pez de~Haro, and A.~Santos.
\newblock Erratum: ``equation of state of a seven-dimensional hard-sphere
  fluid. {P}ercus-{Y}evick theory and molecular-dynamics simulations''.
\newblock {\em J. Chem. Phys.}, 125(21):219903, 2006.

\bibitem{bishop:034506}
M.~Bishop, N.~Clisby, and P.~A. Whitlock.
\newblock The equation of state of hard hyperspheres in nine dimensions for low
  to moderate densities.
\newblock {\em J. Chem. Phys.}, 128(3):034506, 2008.

\bibitem{adda-bedia:184508}
M.~Adda-Bedia, E.~Katzav, and D.~Vella.
\newblock Solution of the {P}ercus-{Y}evick equation for hard disks.
\newblock {\em J. Chem. Phys.}, 128(18):184508, 2008.

\bibitem{adda-bedia:144506}
M.~Adda-Bedia, E.~Katzav, and D.~Vella.
\newblock Solution of the percus--yevick equation for hard hyperspheres in even
  dimensions.
\newblock {\em J. Chem. Phys.}, 129(14):144506, 2008.

\bibitem{Lishchuk2006}
S.V. Lishchuk.
\newblock Equation of state of the hard-disk fluid on a sphere from
  percus-yevick equation.
\newblock {\em Physica A}, 369(2):266 -- 274, 2006.

\bibitem{PhysRevB.33.499}
J.~M. Caillol and D.~Levesque.
\newblock Low-density phase diagram of the two-dimensional {C}oulomb gas.
\newblock {\em Phys. Rev. B}, 33(1):499--509, 1986.

\bibitem{PhysRevB.55.522}
J.~Lidmar and M.~Wallin.
\newblock Monte {C}arlo simulation of a two-dimensional continuum {C}oulomb
  gas.
\newblock {\em Phys. Rev. B}, 55(1):522--530, 1997.

\bibitem{Torres2004}
A.~Torres and G.~T\'{e}llez.
\newblock Finite-size corrections for {C}oulomb systems in the
  {D}ebye-{H}\"{u}ckel regime.
\newblock {\em J. Phys. A: Math. Gen.}, 37(6):2121--2137, 2004.

\bibitem{Hastings1998}
M.B. Hastings.
\newblock Non-hermitian fermion mapping for one-component plasma.
\newblock {\em J. Stat. Phys.}, 90(1):311--326, 1998.

\bibitem{Hansen1983}
J.P. Hansen and P.~Viot.
\newblock Pair correlations and internal energy of the two-dimensional
  {C}oulomb gas.
\newblock {\em Phys. Lett. A}, 95(3-4):155--158, 1983.

\bibitem{samaj2000}
L.~Samaj and I.~Travenec.
\newblock Thermodynamic properties of the two-dimensional two-component plasma.
\newblock {\em Journal of Statistical Physics}, 101(3):713--730, November 2000.

\bibitem{Samaj2001}
L.~Samaj.
\newblock Universal finite-size effects in the two-dimensional asymmetric
  coulomb gas on a sphere.
\newblock {\em Physica A: Statistical Mechanics and its Applications},
  297(1-2):142 -- 156, 2001.

\bibitem{Samaj2002}
L.~Samaj and B.~Jancovici.
\newblock Large-distance behavior of particle correlations in the
  two-dimensional two-component plasma.
\newblock {\em J. Stat. Phys.}, 106:301--321, 2002.

\bibitem{Samaj2002b}
L.~Samaj and B.~Jancovici.
\newblock Density correlations in the two-dimensional coulomb gas.
\newblock {\em J. Stat. Phys.}, 106:323--355, 2002.

\bibitem{GAUDIN1985}
M.~Gaudin.
\newblock Critical isotherm of a lattice plasma.
\newblock {\em J. Phys. France}, 46(7):1027--1042, 1985.

\bibitem{cornu:2444}
F.~Cornu and B.~Jancovici.
\newblock The electrical double layer: A solvable model.
\newblock {\em J. Chem. Phys.}, 90(4):2444--2452, 1989.

\bibitem{tellez:8572}
G.~T\'{e}llez.
\newblock Two-component plasma in a gravitational field.
\newblock {\em J. Chem. Phys.}, 106(20):8572--8578, 1997.

\bibitem{santos:4622}
A.~Santos, M.~L\'{o}pez de~Haro, and S.~Bravo Yuste.
\newblock An accurate and simple equation of state for hard disks.
\newblock {\em J. Chem. Phys.}, 103(11):4622--4625, 1995.

\bibitem{glandt:4503}
E.~D. Glandt and D.~D. Fitts.
\newblock {P}ercus--{Y}evick equation of state for the two-dimensional
  {L}ennard-{J}ones fluid.
\newblock {\em J. Chem. Phys.}, 66(10):4503--4508, 1977.

\bibitem{glandt:4546}
E.~D. Glandt and D.~D. Fitts.
\newblock Hypernetted-chain equation of state for the two-dimensional
  {L}ennard-{J}ones fluid.
\newblock {\em J. Chem. Phys.}, 68(10):4546--4550, 1978.

\bibitem{singh:5463}
R.~R. Singh, K.~S. Pitzer, J.~J. de~Pablo, and J.~M. Prausnitz.
\newblock Monte {C}arlo simulation of phase equilibria for the two-dimensional
  {L}ennard-{J}ones fluid in the {G}ibbs ensemble.
\newblock {\em J. Chem. Phys.}, 92(9):5463--5466, 1990.

\bibitem{Zerah1985}
G.~Zerah.
\newblock An efficient {N}ewton's method for the numerical solution of fluid
  integral equations.
\newblock {\em Journal of Computational Physics}, 61(2):280--285, 1985.

\bibitem{belloni:8080}
L.~Belloni.
\newblock Inability of the hypernetted chain integral equation to exhibit a
  spinodal line.
\newblock {\em The Journal of Chemical Physics}, 98(10):8080--8095, 1993.

\bibitem{Gillan1979}
M.~J. Gillan.
\newblock A new method of solving the liquid structure integral equations.
\newblock {\em Mol. Phys.}, 38(6):1781--1794, 1979.

\bibitem{Goldstein1969}
M.~Goldstein.
\newblock Viscous liquids and the glass transition: {A} potential energy
  barrier picture.
\newblock {\em J. Chem. Phys.}, 51(9):3728, 1969.

\bibitem{Stillinger1995}
F.~H. Stillinger.
\newblock A topographic view of supercooled liquids and glass formation.
\newblock {\em Science}, 267:1935, 1995.

\bibitem{Kirkpatrick1989}
T.~R. Kirkpatrick, D.~Thirumalai, and Peter~G. Wolynes.
\newblock Scaling concepts for the dynamics of viscous liquids near an ideal
  glassy state.
\newblock {\em Phys. Rev. A}, 40(2):1045, 1989.

\bibitem{PhysRevLett.54.1059}
Y.~Singh, J.~P. Stoessel, and P.~G. Wolynes.
\newblock Hard-sphere glass and the density-functional theory of aperiodic
  crystals.
\newblock {\em Phys. Rev. Lett.}, 54(10):1059--1062, 1985.

\bibitem{M'ezard1999}
M.~M{\'e}zard and G.~Parisi.
\newblock Thermodynamics of glasses: A first principles computation.
\newblock {\em Phys. Rev. Lett.}, 82(4):747, 1999.

\bibitem{pz08}
G.~Parisi and F.~Zamponi.
\newblock Mean field theory of the glass transition and jamming of hard
  spheres.
\newblock arXiv:0802.2180v2.

\bibitem{gonzalez97}
B.~J. Gonzalez and E.~R. Negrin.
\newblock Mehler-{F}ock transforms of generalized functions via the method of
  adjoints.
\newblock {\em Proc. Amer. Math. Soc.}, 125:3243--3253, 1997.

\bibitem{Brader2006}
J.~Brader.
\newblock Solution of the {O}rnstein-{Z}ernike equation in the critical region.
\newblock {\em Int.. J. Thermophys.}, 27(2):394--412, 2006.

\end{thebibliography}

\end{document}